\documentclass[%
10pt,
twocolumn,
 amsmath,amssymb,
 aps,
 superscriptaddress,
pra,
floatfix
]{revtex4-2}
\usepackage{graphicx}
\usepackage{dcolumn}  
\usepackage{bm}       
\usepackage{svg}      
\usepackage{upgreek}  
\usepackage{amssymb}  
\usepackage{makecell} 
\usepackage{xcolor}   
\usepackage[normalem]{ulem} 
\usepackage{todonotes} 
\usepackage{braket}   
\usepackage{url}      
\usepackage{hyperref}
\usepackage[capitalise]{cleveref} 

\definecolor{DarkGreen}{rgb}{0, 0.5, 0}

\usepackage{soul}
\setstcolor{red} 

\makeatletter
\newsavebox{\@brx}
\newcommand{\llangle}[1][]{\savebox{\@brx}{\(\m@th{#1\langle}\)}%
  \mathopen{\copy\@brx\kern-0.5\wd\@brx\usebox{\@brx}}}
\newcommand{\rrangle}[1][]{\savebox{\@brx}{\(\m@th{#1\rangle}\)}%
  \mathclose{\copy\@brx\kern-0.5\wd\@brx\usebox{\@brx}}}
\makeatother

\newboolean{showkeys}
\setboolean{showkeys}{false}
\ifthenelse{\boolean{showkeys}}{\usepackage[notref,notcite]{showkeys}} 
{}

\usepackage{scalerel}
\usepackage{stackengine}

\newcommand{\approptoinn}[2]{\mathrel{\vcenter{
  \offinterlineskip\halign{\hfil$##$\cr
    #1\propto\cr\noalign{\kern2pt}#1\sim\cr\noalign{\kern-2pt}}}}}

\begin{document}

\title{Scaling the Variational Quantum Eigensolver for Dynamic Portfolio Optimization}

\author{Á. Nodar}
\affiliation{Global Data Quantum, Gran Vía de Don Diego López de Haro, 1, 48001 Bilbo, Bizkaia, Spain}
\author{I. De León}
\affiliation{Global Data Quantum, Gran Vía de Don Diego López de Haro, 1, 48001 Bilbo, Bizkaia, Spain}
\author{D. Arias}
\affiliation{Global Data Quantum, Gran Vía de Don Diego López de Haro, 1, 48001 Bilbo, Bizkaia, Spain}
\affiliation{Universidad de Deusto, Avda. de las Universidades, 24, 48007, Bilbo, Bizkaia, Spain}
\author{E. Mamedaliev}
\affiliation{Global Data Quantum, Gran Vía de Don Diego López de Haro, 1, 48001 Bilbo, Bizkaia, Spain}
\author{M. E. Molina}
\affiliation{Global Data Quantum, Gran Vía de Don Diego López de Haro, 1, 48001 Bilbo, Bizkaia, Spain}
\author{M. Martín-Cordero}
\affiliation{Global Data Quantum, Gran Vía de Don Diego López de Haro, 1, 48001 Bilbo, Bizkaia, Spain}
\author{S. Hernández-Santana}
\affiliation{BBVA Quantum, Calle Azul 4, 28050, Madrid, Spain}
\author{P. Serrano}
\affiliation{BBVA Quantum, Calle Azul 4, 28050, Madrid, Spain}
\author{M. Arranz}
\affiliation{Global Data Quantum, Gran Vía de Don Diego López de Haro, 1, 48001 Bilbo, Bizkaia, Spain}
\author{O. Mentxaka}
\affiliation{Global Data Quantum, Gran Vía de Don Diego López de Haro, 1, 48001 Bilbo, Bizkaia, Spain}
\author{V. García}
\affiliation{Lantik, Sabino Arana Etorbidea, 44, 48012 Bilbo, Bizkaia, Spain}
\author{G. Carrascal}
\affiliation{IBM Spain, Plaza Pablo Ruiz Picasso 11, 28020, Madrid, Spain}
\author{A. Retolaza}
\affiliation{BBVA Quantum, Calle Azul 4, 28050, Madrid, Spain}
\author{I. Posadillo}
\affiliation{Global Data Quantum, Gran Vía de Don Diego López de Haro, 1, 48001 Bilbo, Bizkaia, Spain}

\date{\today}

\begin{abstract}
This work explores the potential of the Variational Quantum Eigensolver in solving Dynamic Portfolio Optimization problems surpassing the 100 qubit utility frontier. We systematically analyze how to scale this strategy in complexity and size, from 6 to 112 qubits, by testing different combinations of ansatz and optimizer on a real Quantum Processing Unit. We achieve best results by using a combination of a Differential Evolution classical optimizer and an ansatz circuit tailored to both the problem and the properties of the Quantum Processing Unit.
\end{abstract}

\maketitle

\section{Introduction} \label{sec:Intro}

Gate-based Quantum Processing Units (QPUs) with more than 100 qubits are capable of performing computations at a scale beyond the capabilities of certain classical computing methods. The recent availability of these devices to the general public~\cite{Release100QPU} signaled the beginning of a new era for quantum utility, as these larger QPUs offer great potential in a wide range of applications in realistic scenarios~\cite{kim2023evidence, Flöther_2023, 10628295, Rawat_Mehra_Bist_Yusup_Sanjaya_2022, hassija2020forthcoming, paudel2022quantum}.

The Dynamic Portfolio Optimization (DPO) problem stands as a compelling example in which quantum computing has a potential advantage over classical approaches~\cite{mugel2022dynamic, carrascal2023backtesting, rebentrost2024quantum, Buonaiuto2023, wang2024improving}. Given a set of assets, the DPO aims to find optimal trades in a series of rebalancing times in order to find the best trading trajectory over certain time period. Traditional methods, based on classical algorithms, often struggle to perform DPO with large-scale data~\cite{michaud2008efficient, gunjan2023brief, meher2024risk}.

Despite the current challenges of quantum computing, such as quantum noise, decoherence, and scalability, faced by practical implementation in current Noise Intermediate-Scale Quantum (NISQ) era devices, recent works have demonstrated the potential of quantum computers to address DPO problems~\cite{mugel2022dynamic, xu2023dynamic,Buonaiuto2023}. Yet, it is important to note that, to the best of our knowledge, there is no study that uses real gate-based QPUs with more than a 100 qubits for this purpose.

In this work, we use the Variational Quantum Eigensolver (VQE) algorithm~\cite{peruzzo2014variational, bharti2022noisy} to address the DPO problem on real gate-based QPUs~\cite{Farrell2024}. Building on this approach, we explore the utility of VQE in addressing large-scale optimization problems. To overcome the 100 qubit frontier~\cite{kim2023evidence}, we systematically enhance the most challenging aspects of the scalability of the VQE. Furthermore, we emphasize that our aim in this work is to demonstrate the application of the VQE in these large problems considering only the raw results of the algorithm. That is, we do not apply any quantum error mitigation techniques in order to check the robustness of this method. 

We provide a methodological dissection on scaling up the VQE approach to larger QPUs and larger DPO problem sizes. In particular, we focus on two key technical aspects: the development of an ansatz circuit tailored to the QPU (we use the IBM Torino QPU with 133 superconducting qubits~\cite{IBM_debut_heron}) and the stochastic-based optimizer, Differential Evolution (DE)~\cite{scipyoptimizedifferential_evolution_nodate, carrascal2024differential, storn1997differential, robert2019resource}. The tailored ansatz is specifically designed to capture the structure of the problem to find a solution while being efficient on the QPU architecture. On the other hand, the DE algorithm enhances the optimization process by making it more robust to quantum noise~\cite{PhysRevA.108.032409, article, Pellow_Jarman_2021, diez2023multiobjective} and better addressing barren plateau scenarios~\cite{Nadori:2024twv} than gradient-based optimizers. 

\begin{figure}[t]
    \centering
    \includegraphics[width=.9\columnwidth]{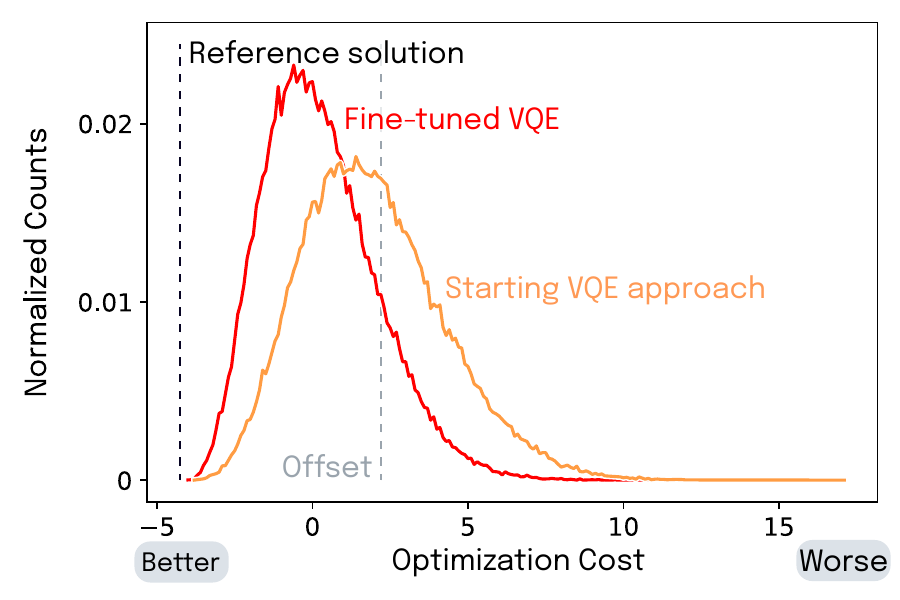}
    \caption{Optimization cost distributions obtained for the starting VQE approach (orange line), and the fine-tuned VQE implementation (red line) that we develop in this work. These distributions are obtained from QPU calculations in the IBM Torino solving the 112 qubits (XXL sized) DPO problem presented in Section~\ref{sec:DPortfolio}. For reference we also give the solution obtained using the Gurobi classical optimizer (black dashed line) and the offset of the problem (average of the random distribution, gray dashed line). We add the optimization costs labels ``better'' and ``worse'' to reinforce that we address the DPO as a minimization problem. To obtain each cost distribution we account for how many times we found investment trajectories (quantum states) with the same associated cost (rounded to the second decimal value).}
    \label{fig:MainResult}
\end{figure}

Figure~\ref{fig:MainResult} summarizes the enhancements in the VQE implementation achieved in this work. The red and orange lines show the results of two VQE implementations for addressing a large DPO problem requiring 112 qubits. The orange line represents our starting VQE approach, reproducing an implementation of the VQE algorithm that has been demonstrated to solve small DPO problems~\cite{mugel2022dynamic}, while the red line corresponds to the best fine-tuned implementation of the VQE that we reach here---the latter includes the tailored ansatz and the DE optimizer. We observe a significant improvement of the VQE results between the starting VQE approach and the fine-tuned implementation. The latter pushes the resulting distribution further away from the offset (average of a random distribution, gray dashed line) towards the optimal reference solution (obtained using the classical Gurobi optimizer, black dashed line).

This work is structured as follows: we first formulate and give the charactersitics of the DPO problems solved in this work in Section~\ref{sec:DPortfolio}. Next, in Section~\ref{sec:ClassBench}, we present a classical benchmark for these DPO problems. In Section~\ref{sec:VQE} we review the VQE algorithm. In Section~\ref{sec:PortfolioOptimization} we introduce the different VQE implementations explored in this work, and their performance in solving DPO problems with increasing size.


\section{Dynamic Portfolio Optimization}
\label{sec:DPortfolio}

In this section, we review the DPO problem. Subsection \ref{sec:Formulation} introduces the formulation of this problem, and Subsection~\ref{sec:QUBO} details the mapping of this formulation into a Quadratic Binary Unconstrained Optimization (QUBO) problem, which can be solved with standard quantum optimizers. As highlighted in the Introduction, our goal is to demonstrate the potential of Quantum Computing in solving the DPO problem. We follow the DPO formulation presented in Refs.~\cite{mugel2022dynamic, rosenberg2016solving}, prioritizing illustrative value over the representation of a fully realistic business scenario.

\subsection{Formulation of the Dynamic Portfolio Optimization problem}
\label{sec:Formulation}

Building upon the foundational work of Harry Markowitz in Modern Portfolio Theory~\cite{markowits1952portfolio}, portfolio optimization has become a widely studied mathematical problem in finance. It consists in finding the investment trajectory $\bm{\Omega}$ that maximizes the return of the portfolio, $F(\bm{\Omega})$, while minimizing the risk associated with it, $R(\bm{\Omega})$. The performance of $\bm{\Omega}$ is commonly evaluated by the Sharpe ratio~\cite{sharpe1966mutual},
\begin{equation}
\label{eq:sharperatio}
    S(\bm{\Omega}) = \frac{F(\bm{\Omega})}{\sqrt{R(\bm{\Omega})}},
\end{equation}
where the investment that maximizes $S(\bm{\Omega})$ (maximizing returns and/or minimizing risks) often results in selecting a mix of assets not perfectly correlated, thereby diversifying the portfolio.

In this paper, we explore DPO, a variant of the standard portfolio optimization problem, where the portfolio (with $N_a$ assets) is rebalanced at certain dates (over $N_t$ dates). Rebalancing the portfolio introduces transaction costs, $C(\bm{\Omega})$, that we also consider in the problem. 

Optimizing a DPO objective function directly based on the Sharpe ratio can be challenging due to the inverse power dependence on $\bm{\Omega}$ in Eq.~\eqref{eq:sharperatio}. Instead, we follow the DPO formulation given in Refs.~\cite{mugel2022dynamic, rosenberg2016solving}, where the optimization objetive function to maximize is:
\begin{equation}
    O(\bm{\Omega}) =
    F(\bm{\Omega}) - 
    \frac{\gamma}{2} R(\bm{\Omega}) -
    C(\bm{\Omega}).
    \label{eq:MainObjective}
\end{equation}
Here, $\bm{\Omega}$ is defined as a collection of investments ${\omega}_{t,a}$, the amount of money (in currency units) associated to the asset $a$ at time $t$. Specifically, we first define $\bm{\omega}_t = (\omega_{t, 0}, \dots, \omega_{t, N_a - 1})$ as the investment across all assets in the portfolio at time $t$. Thereafter, $\bm{\Omega} = \{\bm{\omega}_{0}, \dots , \bm{\omega}_{N_{t}-1}\}$ is defined as the collection of the $\bm{\omega}_t$ investment vectors across the $N_t$ rebalancing times. The discretization in time is achieved by defining a constant time interval between rebalancing events. In this study, we consider $\Delta t = 30$ days (see Appendix~\ref{app:Data}).

$F(\bm{\Omega}), R(\bm{\Omega})$, and $C(\bm{\Omega})$, in Eq.~\eqref{eq:MainObjective} are calculated over a range of $N_a$ assets and across $N_t$ time steps. In the expression above, $\gamma$ is the risk aversion coefficient. We found that (for the DPO problems solved here) $\gamma = 1000$ ensures same order of magnitude for the return and the risk in Eq.~\eqref{eq:MainObjective}\footnote{In our work we chose parameter values that ensure all key aspects of the objective function in Eq.~\eqref{eq:MainObjective} are well represented, in order to capture the full complexity of the problem. For example, we do not chose the value of the risk aversion coefficient $\gamma$ according to industry standards, but we chose a $\gamma$ value that ensures the risk function $R(\bm{\Omega})$ and the expected return function $F(\bm{\Omega})$ have the same influence on the final solution.}. Specifically, the expected return is defined as
\begin{equation}
    F(\bm{\Omega}) = \sum_{t=0}^{N_{t}-1} \bm{\mu}_t^T\bm{\omega}_t,
    \label{eq:return}
\end{equation}
where $\bm{\mu}_t = (\mu_{t, 0}, \dots, \mu_{t, N_a-1})$ is the return vector. The components of $\bm{\mu}_t$ are given by the logarithmic return~\footnote{The expression of the logarithmic expected return we use in Eq.~\eqref{eq:mu} can be traced back to the expected bare return of the investment. Specifically, the $\omega_{t,a}$ money invested in an asset $a$ at time $t$ becomes $\mu^{\text{bare}}_{t, a}\times \omega_{t}$ at the next investment window, with the bare return $\mu^{\text{bare}}_{t, a} = 1 + (P_{t+1,a} - P_{t,a})/P_{t, a}$. In Eq.~\eqref{eq:mu} we take the logarithmic of the bare return, $\mu_{t,a} = \log(\mu^{\text{bare}}_{t, a})$.} of each asset $a$, which are calculated from the changes of their prices at time $t+1$, $P_{t+1,a}$, compared to the previous time price, $P_{t, a}$,
\begin{equation}
    \mu_{t,a} = \log \left( \frac{P_{t+1,a}}{P_{t,a}} \right).
    \label{eq:mu}
\end{equation}
The values of $P_{t,a}$ used in this work can be found in Appendix~\ref{app:Data}.

The risk is given by
\begin{equation}
    R(\bm{\Omega}) =  \sum_{t = 0}^{N_t-1} \bm{\omega}_t^T \Sigma_t \bm{\omega}_t,
    \label{eq:risk}
\end{equation}
where $\Sigma_t$ is the covariance matrix of the logarithmic returns calculated for each time step (see more details on the calculation of $\Sigma_t$ in Appendix~\ref{app:Data}).

Last, transaction costs are calculated as:
\begin{equation}
    C(\bm{\Omega}) = \nu \sum_{t=0}^{N_t-1} |\bm{\omega}_t - \bm{\omega}_{t-1}|,
    \label{eq:TransactionCost}
\end{equation}
where $\nu$ is the transaction fee (for simplicity, considered here constant, $\nu = 1\%$, for all assets and times). $C(\bm{\Omega})$ links different time steps, increasing the complexity of the problem, since it can no longer be solved for each time step independently.

Note that $t=0$ introduces the $\bm{\omega}_{-1}$ term in Eq.~\eqref{eq:TransactionCost}, which corresponds to the initial condition of the problem. We chose to set $\bm{\omega}_{-1}=\bm{0}_{N_{a}}$ as if there is no initial investment.

The objective function for this problem is subjected to the following budget constraints:
\begin{equation}
    \omega_{t,a} \leq K',
    \label{eq: constraint maximum invest asset}
\end{equation}
and
\begin{equation}
    \sum_{a=0}^{N_a-1} \omega_{t,a} = K, \qquad \forall t.
    \label{constraint_original_problem}
\end{equation}
For a given time $t$, we limit the purchase to $K'$ money invested in a single asset and a portfolio budget of $K$ across all assets.

\subsection{Dynamic Portfolio Optimization problem as a Quadratic Unconstrained Binary Optimization problem}
\label{sec:QUBO}

In order to work with the VQE algorithm, we transform the objective function $O(\bm{\Omega})$ in Eq.~\eqref{eq:MainObjective} into a QUBO problem. By definition, a QUBO problem consists of binary variables with terms that are either linear or quadratic, and it must exclude constraints.

Following Ref.~\cite{mugel2022dynamic} we define the QUBO problem with respect to the normalized investment trajectory (normalized by the portfolio budget $K$),
\begin{equation}
    \bm{\tilde{\Omega}}:= \frac{\bm{\Omega}}{K},
    \label{eq:normalized_investment_traj}
\end{equation}
and the normalized investments at asset $a$ and time $t$,
\begin{equation}
    \tilde{\omega}_{t,a}:= \frac{\omega_{t,a}}{K}.
    \label{eq:normalized_investment}
\end{equation}
With this renormalization, Eqs.~\eqref{eq: constraint maximum invest asset} and~\eqref{constraint_original_problem} become
\begin{equation}
    \tilde{\omega}_{t,a} \leq\frac{K'}{K},
    \label{eq: constraint maximum invest asset_normalized}
\end{equation}
and
\begin{equation}
    \sum_{a=0}^{N_a-1} \tilde{\omega}_{t,a} = 1, \qquad \forall t,
    \label{constraint_original_problem_normalized}
\end{equation}
respectively.

For convenience, we also modify the objective function converting it from a maximization problem to a minimization problem, since many quantum optimization algorithms (\textit{e.g.}, adiabatic evolution) are framed towards minimization. Thus, $O(\bm{\Omega})$ in Eq.~\eqref{eq:MainObjective} becomes:

\begin{equation}
    O(\bm{\Omega}) \to O_\text{min}(\tilde{\bm{\Omega}}) = - F(\tilde{\bm{\Omega}}) + R(\tilde{\bm{\Omega}}) + C(\tilde{\bm{\Omega}}).
    \label{eq:MainObjective_minimation}
\end{equation}

Regarding the QUBO formulation, we first tackle the transaction cost function. Due to the absolute value introduced within the transaction costs in Eq.~\eqref{eq:TransactionCost}, the function is neither quadratic nor linear, so we perform a quadratic approximation of the absolute value, which leads us to
\begin{equation}
    C(\tilde{\bm{\Omega}}) = \nu \lambda \sum_{t=0}^{N_{t}-1} (\tilde{\bm{\omega}}_t - \tilde{\bm{\omega}}_{t-1})^2,
    \label{eq:TransactionCost QUBO}
\end{equation}
where $\lambda$ is a parameter introduced to adjust the quadratic difference between weights at different times, so that the error due to the approximation is minimum. By analytically minimizing this error, we find $\lambda = \sqrt[3]{2}K/K'$.

The problem constraint (described in Eq.~\eqref{constraint_original_problem_normalized}) is encoded as a penalty term in the QUBO formulation~\cite{QUBO_tutorial}. In this case, we introduce a quadratic term modulated by a Lagrange multiplier, $\rho$. This way, the constraint becomes soft, so slight deviations can be tolerated. We found that the value of the Lagrange multiplier $\rho=1$ is a good trade-off to reinforce the restrictions while keeping the original objective function from being ruled by the penalty term~\cite{Buonaiuto2023}.

Up to this point, the optimization objective function we work with is
\begin{align}
Q = \sum_{t=0}^{N_t-1}- \bm{\mu}_t^T \tilde{\bm{\omega}}_t + \frac{\gamma}{2} \tilde{\bm{\omega}}_t^T \Sigma_t \tilde{\bm{\omega}}_t +\nu\lambda (\tilde{\bm{\omega}}_t - \tilde{\bm{\omega}}_{t-1})^2 \nonumber\\+
\rho \left( \sum_{a=0}^{N_a-1} \tilde{\omega}_{t,a} -1\right)^2.
    \label{eq: general expresion QUBO}
\end{align}

The last step in the QUBO formulation is to express the investment at a given time $t$, $\tilde{\bm{\omega}}_{t}=(\tilde{\omega}_{t, 0}, \dots, \tilde{\omega}_{t, N_a - 1})$, as binary variables. We achieve this by performing:
\begin{equation}
    \tilde{\omega}_{t,a} = \frac{1}{K} \sum_{r=0}^{N_r-1} 2^r x_{t,a,r},
    \label{binary_conversion}
\end{equation}
with $x_{t,a,r}$ binary variables, and the subindices $t$ and $a$ representing the time and the asset of the investment, respectively. The subindex $r$ ranges from $0$ to $N_r - 1$, where $N_r$ is the amount of binary variables used to describe the normalized investment ($\tilde{\omega}_{t,a}$) for each asset at each time (increasing $N_r$ enhances the precision level defining the investment). Note that, with this conversion, the maximum money invested per asset ($K'$ in Eq.~\eqref{eq: constraint maximum invest asset_normalized}) is $K' = 2^{N_r} - 1$.

Last, we convert the QUBO problem in Eq.~\eqref{eq: general expresion QUBO} into an Ising Hamiltonian, $H_\text{Ising}$, since most quantum optimizers (\textit{e.g.}, quantum annealing) focus on finding the minimum eigenvalue of such Hamiltonians (instead of directly operating on the QUBO problem). Thus, we perform a change of variables from the binary variables $x_{t,a,r}$ to Pauli operators $Z_q$,
\begin{equation}
Q\xrightarrow{ x_{t,a,r} \to \frac{1-Z_{q}}{2}} H_\text{Ising},
    \label{eq: from_qubo_to_ising}
\end{equation}
where $q$ index denotes the number of the logical qubit that corresponds to the binary variable $x_{t,a,r}$, following:
\begin{equation}
    q = r + N_r \times a + t\times (N_a \times N_r),
    \label{eq:qubit_xatr_relation}
\end{equation}
with $q\in[0, N_q - 1]$, being $N_q$ the total number of variables (qubits) required to solve the QUBO problem,
\begin{equation}
    N_q = N_t \times N_a \times N_r.
\end{equation}
This way, we have established the relationship between the optimization objective function in Eq.~\eqref{eq:MainObjective} and the qubits in the parameterized circuit.

In order to solve larger DPO problems using VQE, we analyze the scalability of the VQE by addressing six different problem sizes, each requiring an increasing number of variables. In Table~\ref{tab:Sizes} we describe the six size problems we tackle using the VQE algorithm, indicating the size label of each problem (XS, S, M, L, XL and XXL) with their respective number of time steps ($N_t$), assets ($N_a$), and resolution bits ($N_r$), resulting in a different requirement of variables to solve the problem ($N_q$). Each size in the table has its corresponding total investment ($K$). 

\begin{table}[t]
\scalebox{0.9}{
{\renewcommand{\arraystretch}{1.25}
\begin{tabular}{|c|c|c|c|c|c|}
\hline
\multicolumn{1}{|c|}{\textbf{Size}} & \multicolumn{1}{c|}{$\mathbf{N_t}$} & \multicolumn{1}{c|}{$\mathbf{N_a}$} & \multicolumn{1}{c|}{$\mathbf{N_r}$} & \multicolumn{1}{c|}{$\mathbf{N_q}$}& \multicolumn{1}{c|}{$\mathbf{K}$ \textbf{(currency units)}}\\ \hline
\multicolumn{1}{|c|}{XS}   & \multicolumn{1}{c|}{2}          & \multicolumn{1}{c|}{3}          &\multicolumn{1}{c|}1      &{6} &{2}    \\ \hline
\multicolumn{1}{|c|}{S}    & \multicolumn{1}{c|}{5}          & \multicolumn{1}{c|}{4}          &\multicolumn{1}{c|}1      &{20}  &{3}    \\ \hline
\multicolumn{1}{|c|}{M}    & \multicolumn{1}{c|}{7}          & \multicolumn{1}{c|}{4}          &\multicolumn{1}{c|}1      &{28}    &{3}  \\ \hline
\multicolumn{1}{|c|}{L}    & \multicolumn{1}{c|}{4}          & \multicolumn{1}{c|}{7}          &\multicolumn{1}{c|}2      &{56}   &{5}   \\ \hline
\multicolumn{1}{|c|}{XL}   & \multicolumn{1}{c|}{4}          & \multicolumn{1}{c|}{7}          &\multicolumn{1}{c|}3       &{84}  &{12}   \\ \hline
\multicolumn{1}{|c|}{XXL}  & \multicolumn{1}{c|}{4}          & \multicolumn{1}{c|}{7}          &\multicolumn{1}{c|}4   &{112}&{25}
\\ \hline 
\end{tabular}}}
\caption{Number of time steps $N_t$, number of assets $N_a$, number of specification $N_r$, number of total qubits $N_q$, and budget $K$ used for the different DPO problem sizes solved here.}
\label{tab:Sizes}
\end{table}

In the next section, we give an overview of the benchmarks we have used to solve our final QUBO, before diving into the VQE algorithm and the quantum resolution of the problem.

\section{Classical and Quantum-inspired benchmarks}
\label{sec:ClassBench}

In this section, we implement both classical and quantum-inspired methods to solve the QUBO DPO problem as formulated in Section~\ref{sec:Formulation}. This approach allows us to establish a robust baseline for benchmarking the quantum calculations presented in Section~\ref{sec:PortfolioOptimization}. We begin in Subsection~\ref{sec:ClassicalBenchmarks} by solving the QUBO problem using several classical solvers. In Subsection~\ref{sec:QuantumInspiredBenchmark}, we introduce a quantum-inspired method to compare our quantum calculations with an alternative quantum approach executed on classical hardware (rather than a QPU).

\subsection{Classical Benchmarks}
\label{sec:ClassicalBenchmarks}

We first benchmark four classical reference solvers: exhaustive search, Gurobi~\cite{gurobi}, IBM Decision Optimization CPLEX (DOCPLEX)~\cite{docplex}, and GEKKO~\cite{beal2018GEKKO} (details on Appendix~\ref{app:Benchmarks}). Table~\ref{tab:clasicalcostbenchmarks} shows the minimum optimization cost values found with the four classical solvers across the different problem sizes. The exhaustive method only manages to find the exact optimal solution for the XS, S, and M sizes---the only cases fitting within our computational resources (see Appendix~\ref{app:computational_details}). The exact optimal solution is reached by the DOCPLEX and Gurobi method for the sizes implemented for the exhaustive search method. For larger sizes, L, XL, and XXL, Gurobi finds the lowest optimization cost (optimal) solution. Gurobi gives the same results as DOCPLEX for L and XL sizes, but differs in XXL, where Gurobi performs a better optimization.

We observe that GEKKO returns the worst results for this optimization problem, with the lowest performance across all sizes. We want to emphasize that for this specific benchmark we are using the GEKKO functionality to solve QUBO problems, which is not the specialty of the GEKKO optimization suite.

\begin{table}[t]
\centering
\scalebox{0.9}{
{\renewcommand{\arraystretch}{1.25}
\begin{tabular}{|c|c|c|c|c|c|c|}
\hline
\textbf{Optimizer} & \textbf{XS} & \textbf{S} & \textbf{M} & \textbf{L} & \textbf{XL} & \textbf{XXL} \\ 
\hline
Exhaustive          & -1.99              & -5.04              & -7.14              &    -               &       -            &      -             \\ \hline
DOCPLEX               & -1.99              & -5.04              & -7.14              & -4.24              & -4.25              & -4.24              \\ \hline
Gurobi              & -1.99              & -5.04              & -7.14              & -4.24              & -4.25              & -4.25              \\ \hline
GEKKO               & -1.43              & -4.33              & -6.18              & -4.06              & -3.60              & -3.94              \\ \hline
SAE      & -1.99              & -5.04              & -7.09              & -4.09              & -4.13              & -4.13              \\ \hline
\end{tabular}}}
\caption{Minimum optimization cost values found using different optimizers across all problem sizes.}
\label{tab:clasicalcostbenchmarks}

\end{table}

While the main criterion for evaluating the quality of our solution is the optimization cost, we also evaluate the associated Sharpe value for a complementary analysis of our solutions. The Sharpe ratio values associated to the previous optimization cost results are shown in Table~\ref{tab:clasicalsharpebenchmarks} as calculated using Eq.~\eqref{eq:sharperatio}. From the formulation in Section~\ref{sec:Formulation}, minimizing the optimization cost does not directly imply an increase in the Sharpe ratio. By comparing Table~\ref{tab:clasicalcostbenchmarks} and Table~\ref{tab:clasicalsharpebenchmarks}, we observe that lower optimization costs correspond to higher (better) Sharpe ratios, with a few exceptions. For example, we see that, for size XXL, DOCPLEX reaches an optimization cost value higher (worse) than Gurobi but the Sharpe ratio result is higher (better).

Based on these results, we chose Gurobi as the primary classical benchmark, as it consistently achieves the lowest optimization cost and the (overall) best Sharpe ratios across all DPO sizes, as shown in Tables~\ref{tab:clasicalcostbenchmarks} and~\ref{tab:clasicalsharpebenchmarks}.

\begin{figure}[t]
    \centering
    \includegraphics[width=.475\textwidth]{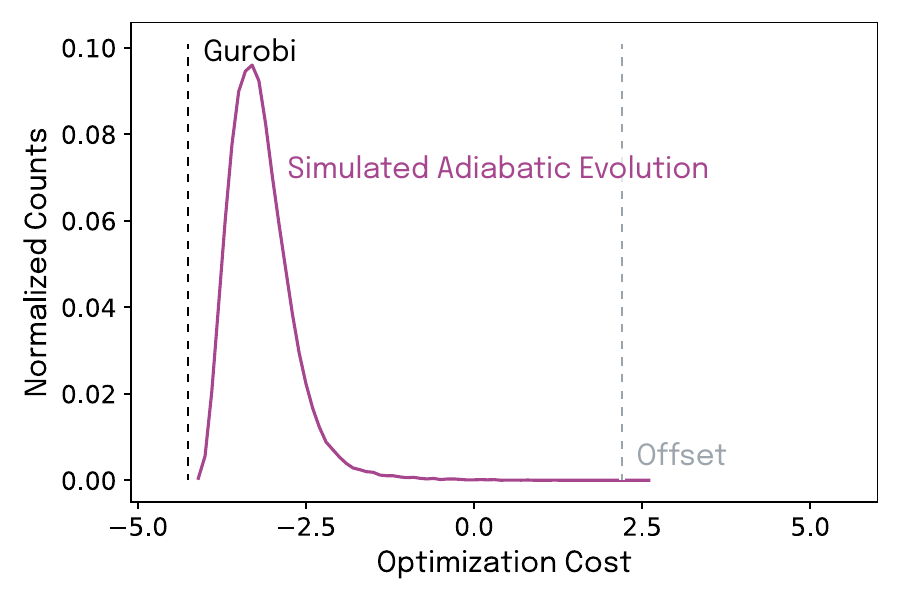}
    \caption{Optimization cost distributions of the solutions obtained for the benchmark problem using SAE. To obtain the optimization cost distribution we account for how many times we found states with the same associated optimization cost (rounded to the second decimal value). We add for comparison the reference values of the offset (corresponding to the average of the random distribution) and the optimization cost obtained by Gurobi.}
    \label{fig:BenchmarkDistributionsXXL}
\end{figure}

\begin{table}[t]
    \centering
    \scalebox{0.9}{
    {\renewcommand{\arraystretch}{1.25}
    \begin{tabular}{|c|c|c|c|c|c|c|}
    \hline
    \textbf{Optimizer} & \textbf{XS} & \textbf{S} & \textbf{M} & \textbf{L} & \textbf{XL} & \textbf{XXL} \\ \hline
    
    Exhaustive          & {5.43}                & {10.67}              & {18.24}               & -                  & -                  & -                  \\ \hline
    DOCPLEX               & {5.43}              & {7.13}               & {16.98}              & {24.55}              & {24.72}               & {24.90}               \\ \hline
    Gurobi              & {5.43}               & {10.67}              & {18.24}              & {24.55}              & {25.02}               & {24.71}               \\ \hline
    GEKKO               & {4.07}               & {4.84}               & {10.31}              & {20.76}              & {12.54}               & {13.00}              \\ \hline
    SAE      & {5.43}                & {10.87}              & {15.90}               & {20.01}               & {18.65}           & {21.61}               \\ \hline
    
    \end{tabular}}}
\caption{Sharpe ratio results of the investment trajectories associated to each minimum optimization cost values shown in Table~\ref{tab:clasicalcostbenchmarks}.}
\label{tab:clasicalsharpebenchmarks}
\end{table}

\subsection{Quantum-Inspired Benchmark}
\label{sec:QuantumInspiredBenchmark}

In addition to our classical benchmarks, we develop a quantum-inspired Tensor Network algorithm using the Python library QUIMB~\cite{Gray2018quimbAP}. We implement a simulated adiabatic evolution (SAE)~\cite{farhi2000quantumcomputationadiabaticevolution} algorithm, following Ref.~\cite{zahedinejad2017combinatorialoptimizationgatemodel}. The SAE allows us to have a benchmark for a quantum strategy run on classical hardware. The details on the implementation of the SAE are beyond the scope of the main text and are described on Appendix~\ref{app:DetailsOnTN}.

We can compare in Tables~\ref{tab:clasicalcostbenchmarks} and~\ref{tab:clasicalsharpebenchmarks} the optimal costs and Sharpe ratios, respectively, achieved by the SAE method across different sized DPO problems. Overall the SAE method gives results close to the ones obtained by Gurobi and DOCPLEX (improving over the GEKKO solution).

In Fig.~\ref{fig:BenchmarkDistributionsXXL} (magenta line) we show the resulting optimization cost distribution obtained from the SAE for the XXL case. This optimization cost distribution corresponds to the number of solutions with the same optimization cost. The SAE method seeks to optimize the distribution by bringing it closer to the Gurobi reference value (dashed black line) and further away from the offset value (average of random distribution, dashed gray line). To asses the bias towards optimal values in the resulting distribution we evaluate the percentage of states between the offset and the reference value of Gurobi. Table~\ref{tab:clasicalcountsbenchmarks} shows these results for the SAE method for all DPO problem sizes. In the XXL case shown in  Fig.~\ref{fig:BenchmarkDistributionsXXL}, the quantum-inspired SAE has effectively biased the probability distribution, with $\sim100\%$ of the states between the offset value and the reference value of Gurobi, demonstrating the capability of the SAE to generate a distribution of optimal solutions.

To conclude this section on benchmarks run over classical hardware, let us point out that Gurobi was the method providing the best results, and thus, we use it for testing the performance of gate-based quantum approaches (in Section~\ref{sec:PortfolioOptimization}, below).

\begin{table}[t]
\centering
\scalebox{0.9}{
{\renewcommand{\arraystretch}{1.25}
\begin{tabular}{|c|c|c|c|c|c|c|}
\hline
\textbf{Optimizer} & \textbf{XS} & \textbf{S} & \textbf{M} & \textbf{L} & \textbf{XL} & \textbf{XXL} \\ \hline
SAE              & 64.06$\%$      & 94.93$\%$      & 99.16$\%$      & 99.78$\%$      & 99.97$\%$      & $\sim$100$\%$      \\ \hline
\end{tabular}}}
\caption{Percentage of measured states with an associated optimization cost below the offset for the SAE optimizer.}
\label{tab:clasicalcountsbenchmarks}
\end{table}


\section{Variational Quantum Eigensolver}
\label{sec:VQE}

We next implement the VQE algorithm to solve the QUBO problem formulated in Eq.~\eqref{eq: general expresion QUBO} in real QPU devices. The VQE algorithm is a leading approach in solving quantum optimization problems, demonstrating significant success on current NISQ-era devices~\cite{gratsea2024onionvqeoptimizationstrategyground, Zhang_2022, bhattacharjee2024}. In this section, we review our implementation of VQE following the scheme in Fig.~\ref{fig:SchemeVQE}. Specifically, our implementation consist on five steps:
\begin{figure}[t]
    \centering
    \includegraphics[width=.475\textwidth]{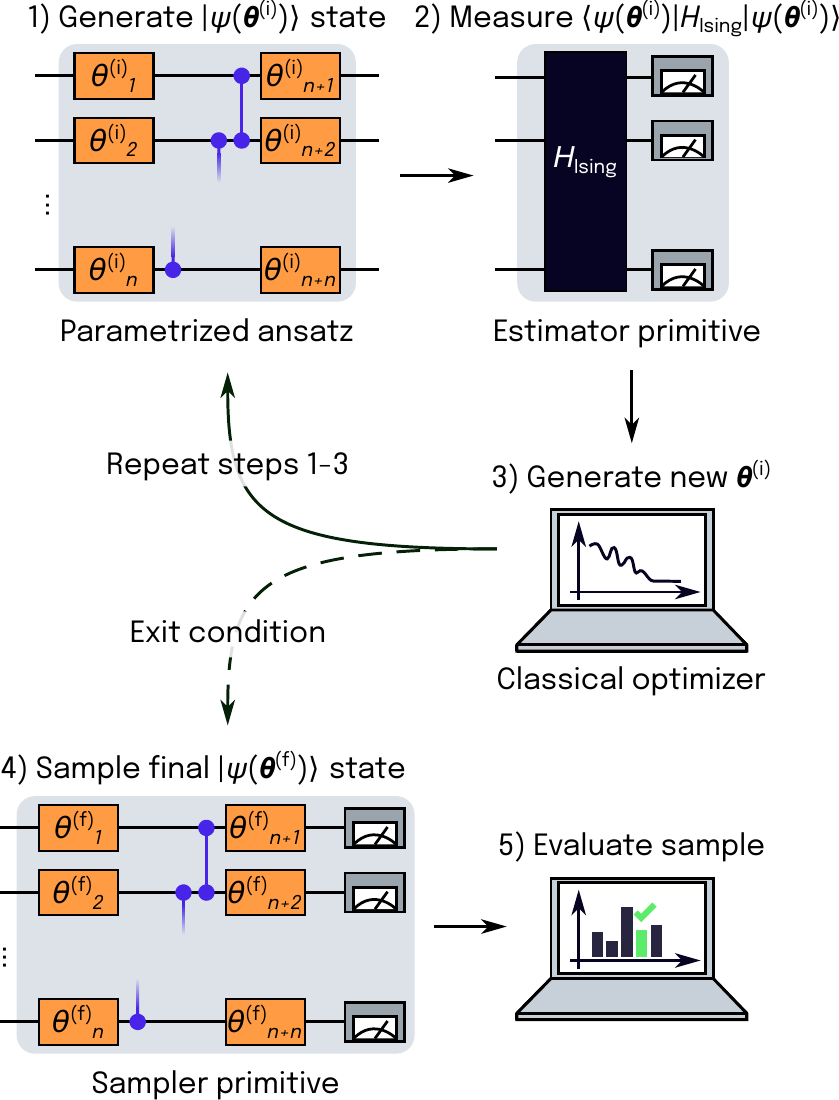}
    \caption{Scheme of the VQE optimization process (described in Section \ref{sec:VQE}): Steps 1-3 comprise generating an ansatz circuit for a given set of parameters $\bm{\theta}^{(i)}$ (step 1), measuring the expected value of the Ising Hamiltonian $H_\text{Ising}$ using the estimator primitive of IBM quantum services (step 2), and optimizing $\bm{\theta}^{(i)}$ with a classical computer (step 3). If the optimal $\bm{\theta}^{(i)}$ is not found in step 3, the process returns to step 1. Once the optimal $\bm{\theta}^{(f)}$ is found, the algorithm proceeds to sample the solution (steps 4 and 5), which involves sampling the ansatz circuit for the optimal $\bm{\theta}^{(f)}$ (step 4, where we use the sampler primitive of IBM quantum services) and identifying the optimal sampled state (step 5).}
    \label{fig:SchemeVQE}
\end{figure}

\begin{enumerate}
    \item The performance of the VQE algorithm heavily relies on choosing a suitable ansatz (parametrized circuit) that assists the quantum optimization. For each $i$-th iteration of the VQE algorithm, we set a $\bm{\theta}^{(i)}$ set of parameters (rotation angles) of the ansatz circuit to generate the quantum state $\ket{\psi(\bm{\theta}^{(i)})}$. Each parameter in $\bm{\theta}^{(i)}$ satisfies $\theta_l^{(i)} \in [-2\pi, 2\pi]$ (see Appendix~\ref{app:ClassicalOpt}).
    
    For the first iteration, $i = 0$, we initialize the parameter vector using random values.
    
    \item We evaluate the expectation value of the Ising Hamiltonian encoding the QUBO problem,
    \begin{equation}
        \braket{\psi(\bm{\theta}^{(i)})|{H}_{\text{Ising}}|\psi(\bm{\theta}^{(i)})}.
    \end{equation}
    Using IBM Quantum QPUs, this can be achieved using the estimator primitive~\cite{qiskit_estimator} (more details in Appendix~\ref{app:computational_details}).
    
    \item We optimize $\braket{\psi(\bm{\theta}^{(i)})|H_\text{Ising}|\psi(\bm{\theta}^{(i)})}$ using a classical optimizer, which is another key aspect on the performance of the VQE algorithm (more details on the classical optimizers in Section~\ref{sec:PortfolioOptimization} and in Appendix~\ref{app:ClassicalOpt}).

    We consider two conditions for exiting the optimization: either the classical optimizer meets the convergence criterion, or the maximum number of iterations is reached. If neither condition is met, we generate a new set of parameters $\bm{\theta}^{(i)}$ and repeat steps 1–3. If one of the exit conditions is satisfied, we define the final $\bm{\theta}^{(i)}$ as the optimized parameters, $\bm{\theta}^{(f)}$, and proceed to the next step.

    \item We sample the state $\ket{\psi(\bm{\theta}^{(f)})}$ using the sampler primitive \cite{qiskit_sampler} (more details in Appendix~\ref{app:computational_details}), obtaining
    \begin{equation}
        \ket{\psi(\bm{\theta}^{(f)})}_{\text{sampled}} = \sum_b A_b \ket{b}, 
    \end{equation} 
    where $A_b$ represents the quasiprobability of measuring the bit string $\ket{b}$.

    \item Finally, we compute the optimization cost associated with each bit string, $c(b) = \braket{b|\hat{H}|b}$, and identify the solution bit string $\ket{b}$ resulting in the smallest $c(b)$ value.
    
\end{enumerate}

In this study, we explore the importance of selecting an appropriate ansatz and classical optimizer for the VQE by evaluating various combinations of classical optimizers and ansatzes. This leads to a comprehensive exploration of large-scale optimization problems.


\section{Quantum Dynamic Portfolio Optimization}
\label{sec:PortfolioOptimization}

As mentioned in the previous section, the performance of the VQE algorithm strongly relies on the chosen classical optimizer and ansatz. In this section, we explore this dependence as we scale the DPO problem from a size requiring 6 qubits to a large scale problem requiring 112 qubits (see Table~\ref{tab:Sizes}) and we systematically solve each problem size for a better comprehension of the scaling for each method. Upscaling the problem not only increases the number of qubits but also the depth of the circuits, leading to decoherence and gate-associated quantum noise. We remark that in our work we do not use any quantum error mitigation. Thus, we aim to find VQE implementations that are robust enough to quantum noise such that their raw implementation gives satisfactory results.

Previous studies discuss that performing the VQE optimization with Hardware Efficient Ansatz (HEA) circuits whose depth grows polynomially with the number of qubits $N_q$, generally leads to barren plateau cases~\cite{Nadori:2024twv, leone_practical_2022, McClean_2018}. That is, as $N_q$ increases, the expected value of the ansatz, $\braket{\psi(\bm{\theta})|H_\text{Ising}|\psi(\bm{\theta})}$, becomes increasingly unresponsive to changes of in the $\bm{\theta}$ parameters. This is the case of the standard HEA ansatzes considered below. Our goal is to explore alternative circuits with smaller depths, and, thus, being less susceptible to barren plateau scenarios.

It is important to note that every circuit must be transpiled~\cite{ibmTranspilerQuantum} before running it on QPU. This transpilation primarily depends on the native gates of the quantum computer and the arrangement of its physical qubits. For example, if two logical qubits that need to be connected are not directly linked on the QPU, SWAP gates are introduced to achieve this entanglement. The transpilation process can significantly increase the depth of each circuit. In this work, we discuss depth after transpilation.

The comparison between methods for all problem sizes in the subsections below is built upon the results in Tables~\ref{tab:QPercentages}-\ref{tab:QSharpe}, where Table~\ref{tab:QPercentages} gives the percentage of measured states with an associated optimization cost below the offset, Table~\ref{tab:QCosts} gives the optimal cost found, and Table~\ref{tab:QSharpe} gives the Sharpe ratio of the investment associated with the optimal cost given in Table~\ref{tab:QCosts}.

\begin{figure}[t]
    \centering
    \includegraphics[width=.475\textwidth]{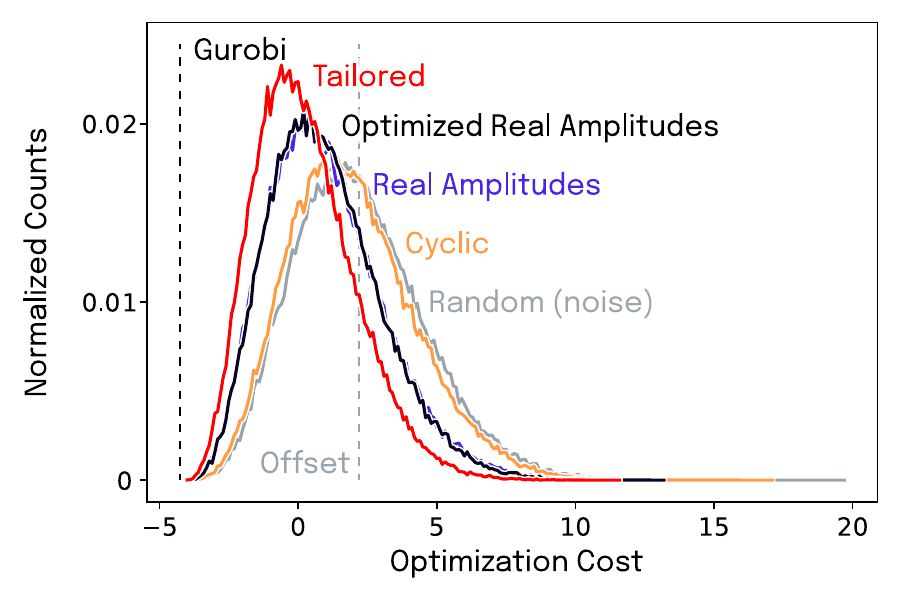}
    \caption{Optimization cost distributions of the VQE solutions to the XXL-sized DPO problem. Each distribution corresponds to the solution of a different ansatz to the VQE process explained in Section~\ref{sec:VQE}. To obtain the optimization cost distribution we account for how many times we found states with the same associated optimization cost (rounded to the second decimal value). We add for comparison the random distribution, the reference values of the offset (corresponding to the average of the random distribution), and the optimization cost obtained by Gurobi (best performing classical method we found in the benchmarks of Section~\ref{sec:ClassBench}).}
    \label{fig:DistributionsXXL}
\end{figure}

We specially focus on the performance of each implementation for the XXL sized DPO problem. In Fig.~\ref{fig:DistributionsXXL}, we show the optimization cost distribution for each VQE implementation discussed in this section. For reference we include the offset value (dashed gray line) and the Gurobi optimization cost (dashed black line). To asses the influence of the quantum noise in this results we include the results of a random sampling in Table~\ref{tab:QPercentages} and Fig.~\ref{fig:DistributionsXXL} (gray line). The random sampling reproduces a purely noisy quantum solution. Since our VQE results are obtained without any quantum error mitigation, the random distribution gives us an additional baseline for comparison.

\begin{figure}[t]
    \centering
    \includegraphics[width=.475\textwidth]{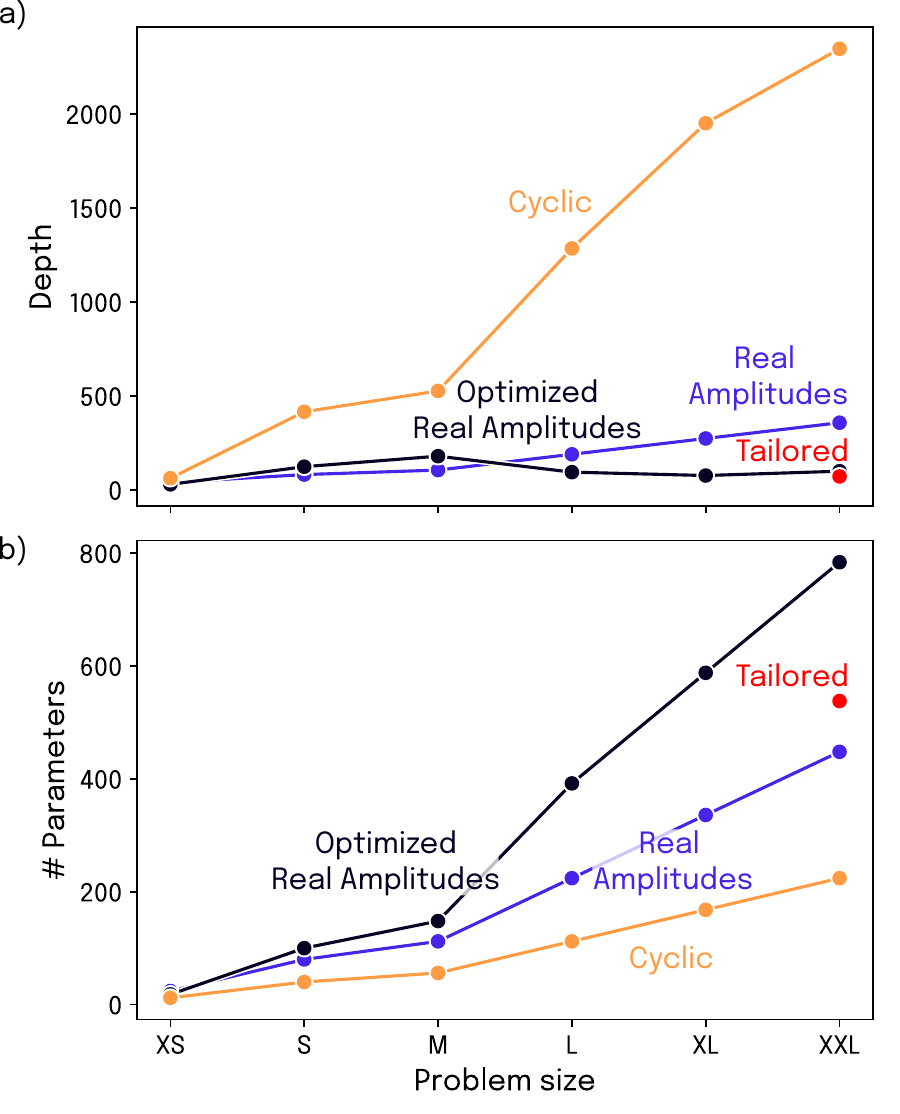}
    \caption{Depth and number of parameters for all ansatz considered in section~\ref{sec:PortfolioOptimization}. Panel a) shows the changes in depths (after transpilation for the IBM Torino QPU) for all the ansatzes as we increase the DPO problem size. Panel b) shows the increase in the number of parameters of each ansatz for the increasing problem size.}
    \label{fig:depth_parameters}
\end{figure}

Additionally, to provide a consistent framework for evaluating the characteristics of each ansatz circuit discussed below, we present the circuit depth in Fig.\ref{fig:depth_parameters}a and the number of parameters to optimize in Fig.\ref{fig:depth_parameters}b.

\begin{table}[t]
\centering
\scalebox{0.9}{
{\renewcommand{\arraystretch}{1.25}
\begin{tabular}{|c|c|c|c|c|c|c|}
\hline
\makecell{\textbf{Ansatz} \\ \textbf{(optimizer)}} & \textbf{XS} & \textbf{S} & \textbf{M} & \textbf{L} & \textbf{XL} & \textbf{XXL} \\ \hline
Cyclic (CG)   & 64.91$\%$ & 51.51$\%$ & 51.36$\%$ & 58.45\% & {60.59$\%$} & 57.95$\%$ \\ \hline
\makecell{Real \\ Amplitudes \\ (DE)} & 64.06$\%$ & 75.36$\%$ & 77.09$\%$ & 73.97$\%$ & 68.92$\%$ & 74.10$\%$ \\ \hline
\makecell{Optimized Real \\ Amplitudes \\ (DE)} & 63.16$\%$ & 76.42$\%$ & 74.88\% & {80.67$\%$} & 74.88 $\%$ & 74.75$\%$\\ \hline
Tailored (DE) & - & - & - & - & - & 85.47$\%$ \\ \hline
\multicolumn{7}{c}{\vspace{-0.3cm}} \\ \hline
Random         & 64.06$\%$    & 54.30$\%$      & 53.79$\%$     & 54.50$\%$      & 54.08$\%$      & 53.60$\%$      \\ \hline
\end{tabular}}
}
\caption{Percentage of measured states with an associated optimization cost below the offset for different combinations of ansatz and classical optimizers across all problem sizes. The combinations are cyclic ansatz with CG and RA, ORA and tailored ansatzes with DE, as indicated in the first column, in parenthesis. For comparison we add the percentages associated with the random (noise) distribution in Fig.~\ref{fig:DistributionsXXL}.}
\label{tab:QPercentages}
\end{table}

\begin{table}[t]
\centering
\scalebox{0.9}{
{\renewcommand{\arraystretch}{1.25}
\begin{tabular}{|c|c|c|c|c|c|c|}
\hline
\textbf{Ansatz (optimizer)} & \textbf{XS} & \textbf{S} & \textbf{M} & \textbf{L} & \textbf{XL} & \textbf{XXL} \\ \hline
Cyclic (CG)   & -1.99 & -4.98 & -7.00 & -3.92 & {-3.94} & -3.76 \\ \hline
\makecell{Real \\ Amplitudes \\ (DE)} & -1.99 & -5.03 & -7.04 & {-4.01} & {-3.95}  & -4.04 \\ \hline
\makecell{Optimized Real \\ Amplitudes \\ (DE)} & -1.99 & -5.04 & -7.05 & {-4.08} & -3.91 & -3.90 \\ \hline
Tailored (DE) & - & - & - & - & - & -4.00 \\ \hline
\end{tabular}}}
\caption{Minimum optimization cost values found using different combinations of ansatz and classical optimizers across all problem sizes. The combinations are cyclic ansatz with CG and RA, ORA and tailored ansatzes with DE, as indicated in the first column, in parenthesis}.
\label{tab:QCosts}
\end{table}

\begin{table}[t]
\centering
\scalebox{0.9}{
{\renewcommand{\arraystretch}{1.25}
\begin{tabular}{|c|c|c|c|c|c|c|}
\hline
\textbf{Ansatz (optimizer)} & \textbf{XS} & \textbf{S} & \textbf{M} & \textbf{L} & \textbf{XL} & \textbf{XXL} \\ \hline
Cyclic (CG)   & {5.43} & {9.02} & {11.60} & {9.35} & {17.67} & {8.30} \\ \hline
\makecell{Real \\ Amplitudes \\ (DE)} & {5.43} & {10.74} & {13.88}& {16.67} & {16.00} & {13.39} \\ \hline
\makecell{Optimized Real \\ Amplitudes \\ (DE)} & {5.43} & {10.67} & {14.40} & {17.43} & {14.16} & {13.18} \\ \hline
Tailored (DE) & - & - & - & - & - & {15.67} \\ \hline
\end{tabular}}}
\caption{Sharpe ratio results of the investment trajectories associated to each minimum optimization cost values shown in Table~\ref{tab:QCosts}. The combinations are cyclic ansatz with CG and RA, ORA and tailored ansatzes with DE, as indicated in the first column, in parenthesis.}
\label{tab:QSharpe}
\end{table}

\subsection{Cyclic ansatz, Conjugate Gradient optimizer}
\label{sec:cyclicCG}

We first reproduce the VQE using the same methodology introduced for the DPO problem in Ref.~\cite{mugel2022dynamic}. In the reference, the authors implemented the VQE algorithm using an ansatz with a cyclic entanglement scheme and a classical Conjugate Gradient (CG) optimizer~\cite{scipy_minimizeme_cg}.

\begin{figure}[t]
    \centering
    \includegraphics[width=.475\textwidth]{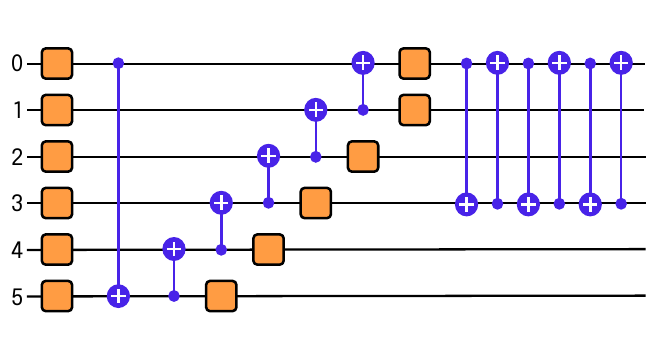}
    \caption{Scheme of cyclic ansatz for 6 qubits (size XS). Each horizontal line represents a logical qubit through the execution time. The orange boxes are rotation gates and the blue circles connected to vertical lines are $\text{C-NOT}$ gates. This circuit represents 3 different assets ($N_a = 3$) in the two studied moments in time ($N_t = 2$). In this case, there is one asset per qubit since the resolution is equal to one ($N_r = 1$).}
    \label{fig:CyclicAnsatz}
\end{figure}

The cyclic ansatz initially proposed on Ref.~\cite{cyclic_ansatz} prepares strongly entangled quantum states. This ansatz is constructed by repetitions of entangling blocks. Each of these blocks consists of a layer of single qubit rotation gates, followed by a sequence of $\text{C-NOT}$ gates arranged according to a specific rule. The number of $\text{C-NOT}$ gates is given by $N_q/\text{gcd}(N_q,d)$ (gcd stands for greatest common divisor), with $N_q$ being the number of qubits of the circuit and $d$ the range modifying the distance of $\text{C-NOT}$ gates~\cite{cyclic_ansatz}. For $j\in[1,\dots, N_q]$ the $j$-th  $\text{C-NOT}$ gate of a block has a target qubit $q_t(j)=d[(N_q - j)-1] \text{ mod } N_q$, and a control qubit $q_c(j)=d(N_q - j) \text{ mod } N_q$. We have used two consecutive blocks of range $d=1$ and range $d=3$, respectively. Fig.~\ref{fig:CyclicAnsatz} represents the scheme of this ansatz for the XS size. However, we note that for the XS size, the global structure of this ansatz is not readily discernible (we have also included an 8-qubit example for method clarification in Appendix~\ref{app:AnsatzesSpec:Cyclic}). 

This entangling structure significantly increases the circuit depth and the number of two-qubit gates (another common source of noise) with each growing step size. Fig.~\ref{fig:depth_parameters}a illustrates the increase in circuit depth of the cyclic ansatz (orange line) as the DPO problem size grows. As shown, the depth of the cyclic ansatz exceeds 1000 for sizes L, XL, and XXL, which is excessive for current NISQ era devices.

Following the workflow in Ref.~\cite{mugel2022dynamic}, we use the CG for the VQE with this ansatz, which is an iterative method for solving systems of linear equations. Nevertheless, as it strongly relies on gradients being found (which can be a problem due to quantum noise and barren plateaus, as mentioned earlier), it has some limitations when it comes to a large set of variables. Specifically, we find that for all sizes the CG achieves convergence (see Appendix~\ref{app:ClassicalOpt}) towards optimization costs very close to the offset of the problem (the average of the random distribution). Thus, we conclude that the standard implementation of the CG is not the best choice for assisting the VQE optimization.

In Fig.~\ref{fig:DistributionsXXL} (orange line, corresponding to orange line in Fig.~\ref{fig:MainResult}) we observe that the cyclic ansatz distribution for the XXL DPO problem is close to the random distribution (gray line). This similarity is expected, given the limitations of the CG optimizer and the ansatz depth for this size (depth is 2347, see Fig.~\ref{fig:depth_parameters}a, orange dot line). We consider that the optimization reached with this VQE implementation is underwhelming, and in the subsections below we explore alternative approaches specifically focusing for this large XXL DPO problem.

\subsection{Real Amplitudes, Differential Evolution optimizer}
\label{sec:EffSU2}

To enhance the VQE implementation and select an ansatz less prone to noise, we use the Real Amplitudes (RA) ansatz~\cite{qiskit_Real_amplitudes}, which prepares states with real-valued amplitudes. Due to its versatility, RA is a common ansatz in chemistry~\cite{kunitsa2024experimentaldemonstrationrobustamplitude, Bauer_2020, anaya2022simulatingmoleculesusingvqe} and machine learning works~\cite{e25070992, sakuma2022resolutionenhancementonedimensionalmolecular,QML_physics}.

\begin{figure}[t]
    \centering
    \includegraphics[width=.475\textwidth]{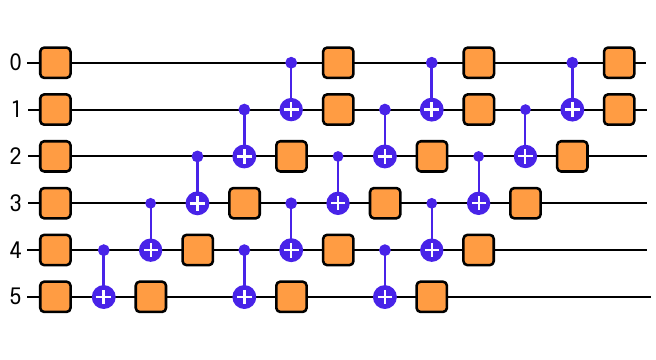}
    \caption{Scheme of RA consisting of 3 layers and entanglement pattern "Inverse linear". In this case, for 
$N_{a}=3$ assets, $N_t=2$ time steps and $N_{r}=1$ specification qubits per asset. Each horizontal line represents a logical qubit throught the execution time. The orange squares in the figure represent $R_Y$ gates and in blue we indicate the $\text{C-NOT}$ gates.}
    \label{fig:RealAmplitudesAnsatz}
\end{figure}

This ansatz consists of a fixed number of layers, each layer being defined by a $R_Y$ rotation on each qubit followed by a $\text{C-NOT}$ gate between each pair of consecutive qubits. In our case, we chose 3 layers with the reverse linear entanglement option---the order in which the rotation gates are applied is inverted, starting from the bottom of the circuit (last qubit), and applying $\text{C-NOT}$ gates upwards, as illustrated in Fig.~\ref{fig:RealAmplitudesAnsatz}. 

In order to optimize the parameters of this ansatz we use DE as the classical optimizer. DE is part of the Genetic Algorithms family~\cite{forrest1996genetic} and is completely gradient-independent, making it more robust to quantum noise~\cite{PhysRevA.108.032409, article, Pellow_Jarman_2021, diez2023multiobjective} and better suited to handle barren plateaus~\cite{Nadori:2024twv}. For details on the DE algorithm implementation, refer to Appendix~\ref{app:ClassicalOpt:DE}.

This is the first scalable VQE implementation in this work that achieves a noticeable difference with the random (noise) distribution for all sizes, with special emphasis on the XXL sized problem (compare blue and gray lines in Fig.~\ref{fig:DistributionsXXL}). Again, this results corresponds to the raw implementation of the VQE, obtained without the use of any quantum error mitigation technique.

Further, for the XXL sized problem, this implementation happens to return the best solution for the VQE, as shown in Table~\ref{tab:QCosts}. However, although the percentage of samples below the offset in Table~\ref{tab:QPercentages} is $74.10\%$, we demonstrate below that other ansatzes could push the distribution closer to the lowest optimization cost. 

In subsections below, we explore two different methods to further exploit the efficiency of the structure of the RA. Specifically, we use $R_Y$ gates and $\text{C-NOT}$ gates connecting first-neighbor qubits. In Subsection~\ref{sec:OptEffSU2}, we first modify the RA ansatz to reflect the relations between variables in our problem (represented by qubits), with the main goal of reducing the overall depth of the circuit. Then, in Subsection~\ref{sec:Tailored}, we explore an ansatz tailored for the XXL problem, which also focuses on reducing the circuit depth by exploiting both the relations between variables in our problem and the structure of the QPU, aiming to push the resulting distribution further towards optimized values.

\subsection{Optimized Real Amplitudes, Differential Evolution optimizer}
\label{sec:OptEffSU2}

\begin{figure}[t]
    \centering
    \includegraphics[width=.5\textwidth]{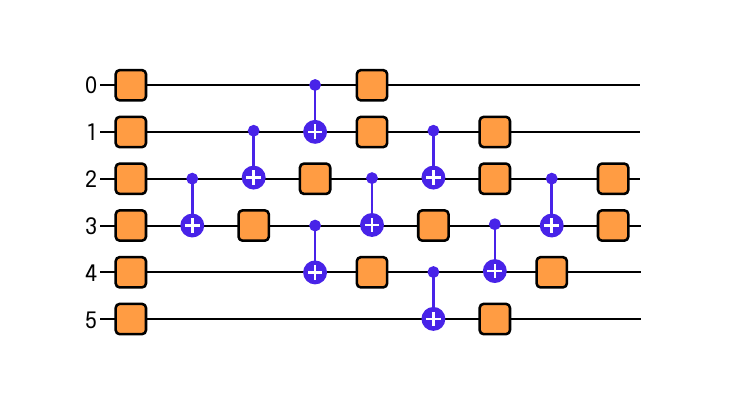}
    \caption{Example of an ORA circuit for a Dynamic Portfolio Optimization problem with $N_a = 3$, $N_t = 2$, and $N_r = 1$. Each horizontal line represents a logical qubit throught the execution time. The orange squares in the figure represent $R_Y$ gates and in blue we indicate the $\text{C-NOT}$ gates.}
    \label{fig:OptRA}
\end{figure}

Next, we propose a systematic circuit structure, the Optimized Real Amplitudes (ORA), based on the RA circuits and the relationship between variables in the QUBO problem. Specifically, we consider RA blocks ranging from the qubits representing the investment $\omega_{t,a}$ of a particular asset $a$ at time $t$ with the qubits representing the investment of the same asset $a$ at time $t+1$, $\omega_{t+1, a}$. This proposal is based on the analysis of the DPO problem formulation in Subsection~\ref{sec:QUBO}, where we observe that cross terms only involve investments of different assets at the same time step, or of the same asset for consecutive time steps, $t$ and $t+1$. 

Figure~\ref{fig:OptRA} shows the ORA ansatz generated for a XS sized DPO problem ($N_a = 3$, $N_t = 2$, and $N_r = 1$). The labels of the qubits are related to the binary variables representing each asset investment following the Eq.~\eqref{eq:qubit_xatr_relation}. As shown in Fig.~\ref{fig:depth_parameters}a, ORA achieves a significant depth reduction compared to RA, especially when the number of qubits per asset exceeds one ($N_r > 1$), \textit{i.e.} for the L, XL, and XXL sizes (see Table~\ref{tab:Sizes}). In particular, for the XXL size, where $N_r = 4$, ORA achieves a depth of 100, while RA has a depth of 358. However, as shown in Fig.~\ref{fig:depth_parameters}b, the reduction in depth comes at the price of an increase in the number of the parameters with the problem size. For example, for the XXL size, the number of parameters of the RA is 448 and for the ORA is 784.

In Table~\ref{tab:QPercentages}, we show that for the L, XL, and XXL sizes (where the ORA results in lower depth than the RA), the ORA outperforms the RA by delivering distributions of solutions with a larger bias towards optimized values. However, while the difference between the ORA and RA is significant (greater than 5$\%$) for the L and XL sizes, both ansatzes yield similar results for the XXL size.

This similarity (for the XXL size) can be further appreciated in Fig.~\ref{fig:DistributionsXXL}, where we observe that the ORA optimization cost distribution (solid black line) is very similar to the one found with the regular RA ansatz (solid blue line). The high number of parameters to optimize (and lack of convergence) of the ORA approach (see Appendix~\ref{app:ClassicalOpt:DE:convergence}), leads us to adopt an approach aimed to reducing the number of parameters in the next subsection.

\subsection{XXL size: Tailored ansatz, Differential Evolution optimizer}
\label{sec:Tailored}

\begin{figure}[t]
    \centering
    \includegraphics[width=.475\textwidth]{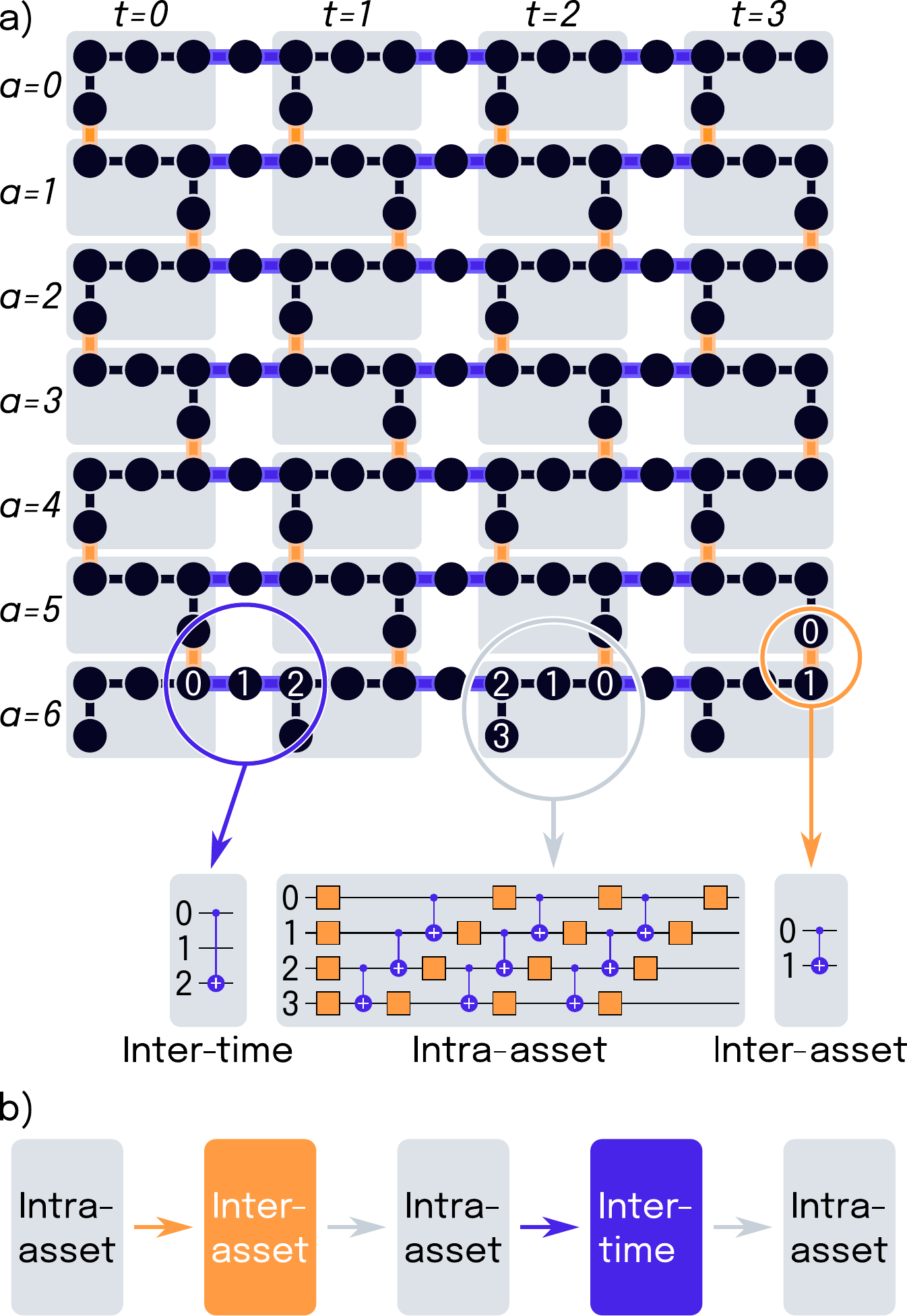}
    \caption{a) Qubit identification of the tailored ansatz from the qubit map of IBM Torino. Gray squares specify qubit subsets of the QPU identified as an asset at a certain instant of time. Orange lines represent the risk interaction between assets. Blue connections represent transaction cost interaction of the same asset at different times. Below the IBM Torino map, we show the circuits for inter-time, intra-asset, and inter-asset transformations, with their orientations. Specifically, each $\text{C-NOT}$ of the inter-time block has its control qubit on the right-most qubit of the IBM Torino map, and each inter-asset $\text{C-NOT}$ block has its control qubit on the upper qubit. In contrast, the intra-asset circuits alternate orientation: for even (odd) asset rows, they start from the upper right (left)-most qubit and proceed to the lower left (right)-most qubit. b) Ansatz circuit scheme. Each step corresponds to a transformation implemented in the quantum circuit.} 
    \label{fig:tailored_7_4_4_identification}
\end{figure}

We take a step further to design an efficient ansatz circuit tailored for both our problem and the IBM Torino QPU. We consider an intermediate approach that could combine the benefits of a HEA with an ansatz tailored to the QUBO problem~\cite{Shen2023, Wang2024, Feulner2022}. For this purpose, we simultaneously consider the relationship between variables in our QUBO problem, their respective mapping to qubits in the quantum circuits, and the connectivity between qubits in the QPU. This strategy helps minimizing the number of $\text{SWAP}$ gates (introduced because of the transpilation) and maximizing the parallel execution of gates, thus reducing circuit depth.

Figure \ref{fig:tailored_7_4_4_identification}a shows the coupling map between the 133 qubits (each represented by a dark blue circle) of the IBM Torino QPU. After evaluating this map against the requirements of the XXL problem—where $N_r = 4$ qubits represent the investment across a portfolio of $N_a = 7$ assets over $N_t = 4$ investment periods---we propose an ansatz circuit with qubits organized into 28 blocks of 4 qubits each, grouped in gray squares in the figure. These blocks are selected using a methodology similar to that in Ref.~\cite{weaving2023contextual}, \textit{i.e.}, forming chains of qubits that interact sequentially with their neighbors. 

For our case, each block corresponds to the investment in a single asset $a$ at a given time $t$, and we consider the existing connections between qubits in the physical QPU by clustering the 4-qubit blocks into vertical and horizontal groups. With this arrangement, the 7 blocks in each of the 4 columns represent the 7 assets at a specific time step.

Thus, according to this block structure, we arrange the gates of the ansatz circuit considering the characteristics of the DPO problem. For each block, the sequence of gates applied (intra-asset circuit) corresponds to a RA circuit with a single repetition, where entanglement is established obeying the physical connectivity of the qubits. Then, for the inter-block gates, we distinguish two types: inter-asset gates, between different assets for the same time step (orange gates in the figure) and inter-time conections, between the same asset for consecutive time steps (blue conexions). Asset gates are made using nearest-neighbor $\text{C-NOT}$ gates between the closest qubits of each asset (the orientation of the control and target gates is indicated in the figure). These $\text{C-NOT}$ gates are executed in parallel and represent changes that affect the entire portfolio at a specific moment. Additionally, inter-time gates between blocks are created using a single $\text{C-NOT}$ gate that connects second-order neighbors. These gates also operate in parallel, with the control qubit again being the one with the lowest index on the QPU map. They represent changes related to how investments in a specific asset evolve over time.

The sequence of operations is executed in parallel following the five steps the scheme in Fig.~\ref{fig:tailored_7_4_4_identification}b: i) we perform all intra-asset circuits, ii) we apply the inter-asset gates, iii) we perform another intra-asset circuits layer, iv) we apply the inter-time C-NOT gates, and, v) we perform the last intra-asset circuit layer. Note that the choice of this ansatz is not unique. We selected this specific ansatz for its simplicity, but various quantum gates could be used for each transformation (asset, risk, and transaction).

The resulting ansatz circuit has a depth of 72, the lowest depth achieved for the XXL size---as shown in Fig.~\ref{fig:depth_parameters}a (red dot). In addition to this significant depth reduction, the tailored ansatz has a lower number of parameters compared to the ORA ansatz, as shown in Fig.~\ref{fig:depth_parameters}b (compare solid black line with red dot, corresponding to ORA and tailored number of parameters, respectively). These lower hardware requirements and lower number of optimization parameters anticipate the potential of this ansatz in order to outperform the results of the other ansatzes.

As observed in Table~\ref{tab:QCosts}, the optimization of this ansatz resulted in a similar cost to the RA and ORA. Additionally, we give in Table~\ref{tab:QSharpe} the Sharpe ratio (see Eq.~\ref{eq:sharperatio}) obtained for the investment with lowest optimization cost, where this tailored ansatz approach achieves the best Sharpe ratio value. 

However, the potential of this ansatz is better appreciated in Fig.~\ref{fig:DistributionsXXL} (red line, corresponding to red line in Fig.~\ref{fig:MainResult}), where the optimization cost distribution presents a significant bias towards the lowest optimization costs, and a smaller standard deviation, meaning it has a trend to favor states (investment trajectories) with optimal associated costs. This confirms the benefits of considering both the QPU and the problem structure when designing the ansatz, enhancing the overall efficiency and accuracy of the quantum algorithm, ensuring that the solution space is explored in a manner aligned with both the hardware capabilities and the inherent features of the problem.
 
This trend and consistency are further reflected in Table~\ref{tab:QPercentages}, where we observe a higher percentage of solutions below the offset. This stability is highly beneficial for real-world applications.

\section{Conclusion and outlooks}
\label{sec:Conclusion}

In this study, we address the DPO problem using the VQE algorithm on a real QPU (we use the IBM Torino QPU). We focus on scaling the VQE algorithm to tackle large DPO problems (requiring more than 100 qubits). With this aim we provide a comprehensive methodological dissection of the various aspects of the regular VQE as we scale the DPO from 6 qubits to 112 qubits. Specifically, we implement different classical optimizers to assist the VQE (Conjugate Gradient and Differential Evolution) as well as various ansatz circuits. These circuits include cyclic ansatz, Real Amplitudes ansatz, Optimized Real Amplitudes ansatz, and an ansatz tailored to both the problem and QPU properties.

We find that, while all combinations of ansatz and classical optimizers successfully solve XS size (6 qubit) DPO problems, larger problems demand precise fine-tuning when configuring the VQE. We observe that in these optimization problems, the DE optimizer is highly effective for the VQE algorithm, achieving convergence in most cases. Most crucially, we find that, for these cases, the most stable and consistent VQE optimization was obtained with an ansatz tailored to both the problem and QPU properties. This highlights the need to not only consider the QPU properties, such as the connectivity and the native quantum gates, but also consider the properties of the optimization problem while designing an ansatz.

Our work has reached the 100-qubit utility frontier using current NISQ-era QPUs on an applied, industry-focused problem. These results set a valuable benchmark for upcoming developments on quantum computing technology and techniques. We have successfully tackled a problem with intrinsic scaling difficulties from 6 to 112 qubits. 
We remark that our results were obtained without any quantum error mitigation technique, which we consider essential to be addressed in future works.


\section*{Acknowledgments}
This work has been conducted within the quantum ecosystem of Bizkaia, under the BIQAIN (Bizkaia Quantum Advanced Industries) initiative, using infrastructures provided by the Government of Bizkaia.

The authors thank Á. Díaz-Fernández, J. Luis-Hita, E. Sánchez-Martínez, J. Fraxanet, and P. Rivero for many useful inputs and valuable discussions.

D. A. acknowledges the partial funding of his doctoral research at the University of Deusto, within the D4K (Deusto for Knowledge) team on applied artificial intelligence and quantum computing technologies, thanks to the support of the Basque Government.

\appendix
\section*{Appendices}


\section{Classical optimizers}
\label{app:ClassicalOpt}

In this study, we perform the VQE optimization using two different classical optimizers (see details of the VQE algorithm in Section~\ref{sec:VQE}). In this appendix we provide the details of the implementation for each classical optimizer.

\subsection*{Conjugate Gradient}

In Subsection~\ref{sec:cyclicCG}, we present the results of reproducing the implementation of the VQE in Ref.~\cite{mugel2022dynamic} to solve the DPO problem. This implementation uses a CG classical optimizer alongside a cyclic ansatz circuit~\cite{schuld2020circuit, mugel2022dynamic}. In particular, we use the conjugate gradient optimizer found in the SciPy minimize function~\cite{scipy_minimizeme_cg}. All hyperparameters were kept at their default values, with the exception of the maximum number of iterations (\texttt{maxiter}) parameter, which was set to 500 iterations.

\subsubsection*{Convergence of the Conjugate Gradient optimizer}

\begin{figure}[t]
    \centering
    \includegraphics[width=.475\textwidth]{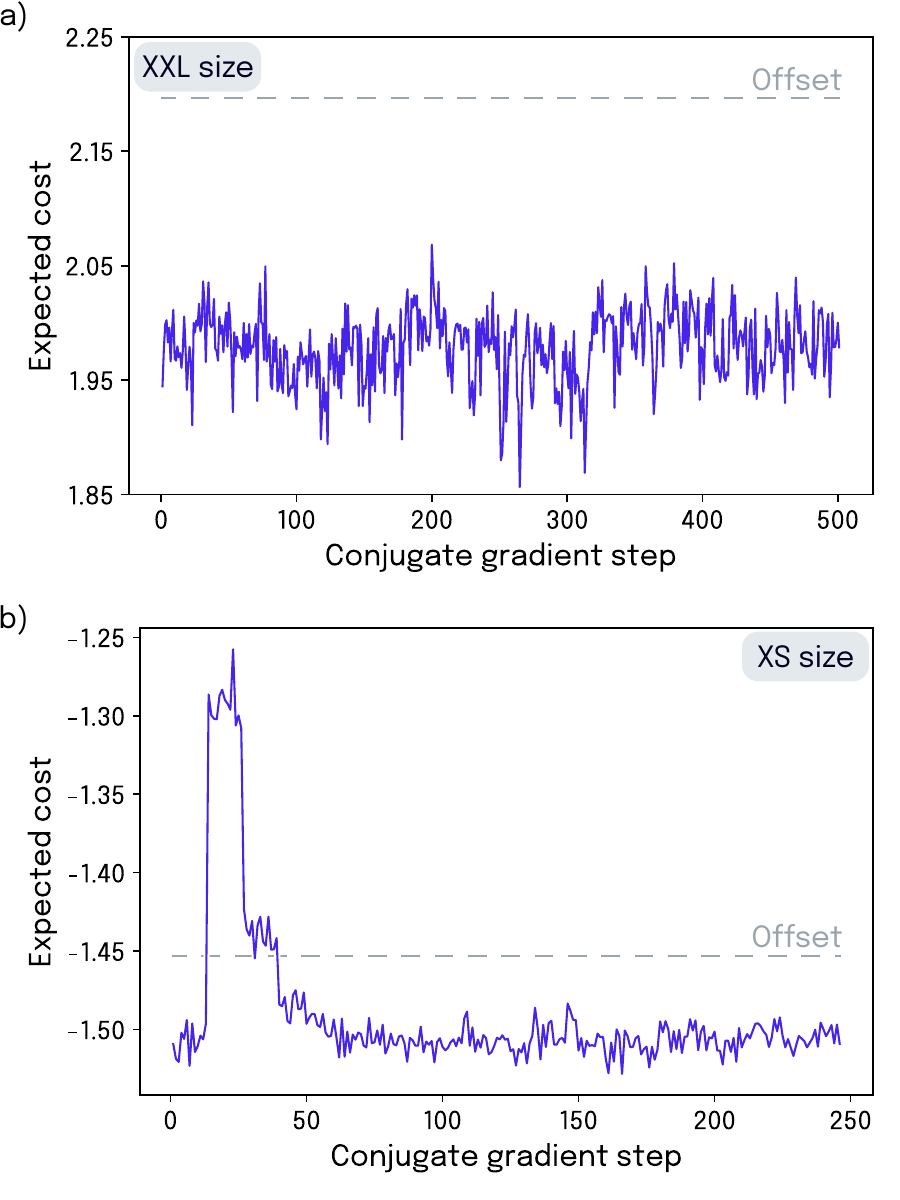}
    \caption{Convergence of the CG method in the VQE optimization of the cyclic ansatz for the DPO problem. Panel a) shows the convergence for the problem size XXL and panel b) for the problem size XS. The blue line shows changes in the expected optimization cost obtained by the estimator process (see Section~\ref{sec:VQE}) at each step of the CG optimizer. The problem offset (average of random a distribution) is indicated by the dashed gray line.}
    \label{fig:ConvergenceCG}
\end{figure}

The combination of the cyclic ansatz and the CG optimization method is not particularly effective for our problem (see Section~\ref{sec:cyclicCG}). We attribute this outcome to two primary factors: the tendency of the CG method to converge to local minima, and the noise introduced by the cyclic ansatz architecture. The latter factor plays a particularly crucial role in DPO problems of sizes S, M, L, XL, and XXL. Both factors affect the convergence of the CG, but in different ways.

For all DPO sizes, the cyclic ansatz results in a distribution very close to a random distribution, a response unsensitive to the changes in the ansatz parameters. As an example, Fig.~\ref{fig:ConvergenceCG}a shows the changes in the expected cost of the cyclic ansatz in the VQE optimization for solving the XXL-sized DPO problem. We observe that the changes are not significant; during the CG optimization, the expected cost fluctuates within a very narrow range $[1.86, 2.07]$, which is a negligible improvement over the offset (corresponding to the expected cost of a random distribution). However, this narrow range of optimization implies convergence. Specifically, for the CG method, we assess convergence by examining the difference in expected cost between the last 100 optimizer steps, which must be less than $2.5\%$ (refer to Table~\ref{tab:ConvergecenCG} for exact values across various problem sizes).

\begin{table}[t]
\scalebox{0.9}{
{\renewcommand{\arraystretch}{1.25}
\begin{tabular}{|c|c|}
\hline
\textbf{Size} & \textbf{Difference in last 100 steps} \\ \hline
XS            &  1.39$\%$ \\ \hline
S             &  0.24$\%$ \\ \hline
M             &  0.11$\%$ \\ \hline
L             &  {0.16$\%$} \\ \hline
XL            &   {6.72 $\%$}                                  \\ \hline
XXL           & 1.01$\%$ \\ \hline
\end{tabular}
}}
\caption{Difference in the expected optimization cost between the last 10 steps of the CG. The difference is giving as a percentage with respect to the value of the mean expected cost of the last step. We give this difference for all the 6 different sizes of the DPO problem considered in Subsection~\ref{sec:cyclicCG}.}
\label{tab:ConvergecenCG}
\end{table}

For the XS-sized problem, the cyclic ansatz achieves a range of expected cost values slightly larger than for other sizes, as shown in Fig.~\ref{fig:ConvergenceCG}b, which illustrates the convergence of the XS-sized DPO problem. However, this increase in range does not imply a successful optimization. In fact, after a sudden large increase in the expected cost around the 20th optimization step, the optimizer manages to find a local minimum and converges to that value at the 246th step.

We emphasize that these results were obtained using the default configuration of the CG implementation in SciPy~\cite{scipy_minimizeme_cg}. We acknowledge that this configuration may be useful for other problems, and that modifying the default settings could potentially improve the CG results.

\subsection*{Differential Evolution}
\label{app:ClassicalOpt:DE}

For the results presented in Subsections~\ref{sec:EffSU2}-\ref{sec:Tailored}, we use a DE optimizer to assist the VQE optimization process. To integrate this optimizer into the VQE workflow, we adapt the differential evolution implementation from SciPy~\cite{scipyoptimizedifferential_evolution_nodate, storn1997differential} modifying the initialization procedure in order to have a better control of the size of the population.

We perform two different initializations depending on the problem size. For the XS problem, each individual of the DE is a list of random parameters. Each element of the random initialization is drawn from the interval $[-2\pi, 2\pi]$. We note that we extend this range to $[-2\pi, 2\pi]$, rather than the typical $[0, 2\pi]$ range used in $R_Y$ gates, in order to facilitate subtraction operations during the crossover phase of the DE algorithm.

For sizes larger than XS, we initiate the DE process by generating a set of initial random parameters for the ansatz exceeding the required population size for DE. For each size larger than XS, we select the best solutions after evaluating approximately $3000$ sets of parameters, and make a selection from them to build the population of the initial DE generation (subsequent generations follow the standard DE rules). This approach implements an elitist strategy in DE~\cite{du2018elitism, zhou2023efficient}, where an initial random search identifies the most promising solutions that serves as a robust foundation for the DE optimization phase.

\begin{table}[t]
\scalebox{0.9}{
{\renewcommand{\arraystretch}{1.25}
\begin{tabular}{|c|c|c|}
\hline
\multicolumn{3}{|c|}{\textbf{Real Amplitudes}}                              \\ \hline
\multicolumn{1}{|c|}{\textbf{Size}} & \multicolumn{1}{c|}{\textbf{Population}} & \multicolumn{1}{|c|}{\textbf{Generations}} \\ \hline
\multicolumn{1}{|c|}{XS}   & \multicolumn{1}{c|}{6}          & \multicolumn{1}{|c|}{50}           \\ \hline
\multicolumn{1}{|c|}{S}    & \multicolumn{1}{c|}{16}          & \multicolumn{1}{|c|}{50}           \\ \hline
\multicolumn{1}{|c|}{M}    & \multicolumn{1}{c|}{24}          & \multicolumn{1}{|c|}{50}           \\ \hline
\multicolumn{1}{|c|}{L}    & \multicolumn{1}{c|}{40}          & \multicolumn{1}{|c|}{50}           \\ \hline
\multicolumn{1}{|c|}{XL}   & \multicolumn{1}{c|}{66}          & \multicolumn{1}{|c|}{50}           \\ \hline
\multicolumn{1}{|c|}{XXL}  & \multicolumn{1}{c|}{92}          & \multicolumn{1}{|c|}{40}         \\ \hline
\end{tabular}}}
\caption{Population size and number of generations used in the DE optimization of the RA ansatzes.}
\label{tab:DiffEv_RealAmp}
\end{table}

\begin{table}[t]
\scalebox{0.9}{
{\renewcommand{\arraystretch}{1.25}
\begin{tabular}{ccc}
\hline
\multicolumn{3}{|c|}{\textbf{Optimized Real Amplitudes}}                              \\ \hline
\multicolumn{1}{|c|}{\textbf{Size}} & \multicolumn{1}{c|}{\textbf{Population}} & \multicolumn{1}{|c|}{\textbf{Generations}} \\ \hline
\multicolumn{1}{|c|}{XS}   & \multicolumn{1}{c|}{6}          & \multicolumn{1}{|c|}{50}           \\ \hline
\multicolumn{1}{|c|}{S}    & \multicolumn{1}{c|}{20}          & \multicolumn{1}{|c|}{50}           \\ \hline
\multicolumn{1}{|c|}{M}    & \multicolumn{1}{c|}{32}          & \multicolumn{1}{|c|}{50}           \\ \hline
\multicolumn{1}{|c|}{L}    & \multicolumn{1}{c|}{80}          & \multicolumn{1}{|c|}{48}           \\ \hline
\multicolumn{1}{|c|}{XL}   & \multicolumn{1}{c|}{120}          & \multicolumn{1}{|c|}{30}           \\ \hline
\multicolumn{1}{|c|}{XXL}  & \multicolumn{1}{c|}{160}          & \multicolumn{1}{|c|}{21}         \\ \hline       
\end{tabular}}}
\caption{Population size and number of generations used in the DE optimization of the ORA ansatzes.}
\label{tab:DiffEv_OptRealAmp}
\end{table}

\begin{table}[t]
\scalebox{0.9}{
{\renewcommand{\arraystretch}{1.25}
\begin{tabular}{ccc}
\hline
\multicolumn{3}{|c|}{\textbf{Tailored}}                              \\ \hline
\multicolumn{1}{|c|}{\textbf{Size}} & \multicolumn{1}{c|}{\textbf{Population}} & \multicolumn{1}{|c|}{\textbf{Generations}} \\ \hline
\multicolumn{1}{|c|}{XXL}   & \multicolumn{1}{c|}{110}          & \multicolumn{1}{|c|}{30}           \\ \hline       
\end{tabular}}}
\caption{Population size and number of generations used in the DE optimization of the tailored ansatz.}
\label{tab:DiffEv_Tailored}
\end{table}

In addition to the modification of the initialization of the DE, we also adapt the standard hyperparameters of the optimizer to our problem. Through prior experimentation with DE applied to VQE (results not shown here), we determined that the optimal hyperparameters for our case, following the notation of SciPy~\cite{scipyoptimizedifferential_evolution_nodate}, are: 
\begin{verbatim}
    strategy = 'best2bin'
    mutation = (0, 0.25)
    recombination = 0.4
\end{verbatim}
All other hyperparameters are set to their default values, except for the population size and number of generations, which vary depending on the ansatz family and system size, as detailed in Tables~\ref{tab:DiffEv_RealAmp}-\ref{tab:DiffEv_Tailored}. We have chosen the values in Tables~\ref{tab:DiffEv_RealAmp}-\ref{tab:DiffEv_Tailored} by fixing a similar number of circuits to evaluate in every ansatz (the amount of circuits to evaluate in the DE optimization corresponds to the product of the population size by the number of generations indicated in the tables). We first choose the population size, being larger for ansatz with more parameters to optimize. Then, we chose the number of generations to have a similar amount of circuits across every ansatz. We have taken this criteria in order to make a fair comparison between the performance of each ansatz being optimized using similar (quantum) computational resources.

\subsubsection*{Convergence of the Differential Evolution optimizer}
\label{app:ClassicalOpt:DE:convergence}

\begin{table}[t]
\scalebox{0.9}{
{\renewcommand{\arraystretch}{1.25}
\begin{tabular}{cccc}
\hline
\multicolumn{4}{|c|}{\textbf{Difference in 10 last generations}}                                    \\ \hline
\multicolumn{1}{|c|}{}    & \multicolumn{3}{c|}{\textbf{Ansatz}}                                            \\ \hline
\multicolumn{1}{|c|}{\textbf{Size}} & \multicolumn{1}{c|}{\begin{tabular}[c]{@{}c@{}} Real \\ Amplitudes\end{tabular}} & \multicolumn{1}{c|}{\begin{tabular}[c]{@{}c@{}}Optimized Real \\ Amplitudes\end{tabular}} & \multicolumn{1}{c|}{Tailored} \\ \hline
\multicolumn{1}{|c|}{XS}  & \multicolumn{1}{c|}{0.09$\%$} & \multicolumn{1}{c|}{3.30$\%$} & \multicolumn{1}{c|}{-} \\ \hline
\multicolumn{1}{|c|}{S}   & \multicolumn{1}{c|}{{0.99 $\%$}} & \multicolumn{1}{c|}{0.10 $\%$} & \multicolumn{1}{c|}{-} \\ \hline
\multicolumn{1}{|c|}{M}   & \multicolumn{1}{c|}{{1.10$\%$}} & \multicolumn{1}{c|}{{0.17$\%$}} & \multicolumn{1}{c|}{-} \\ \hline
\multicolumn{1}{|c|}{L}   & \multicolumn{1}{c|}{{3.31$\%$}} & \multicolumn{1}{c|}{{5.89 $\%$}} & \multicolumn{1}{c|}{-} \\ \hline
\multicolumn{1}{|c|}{XL}  & \multicolumn{1}{c|}{{4.30 $\%$}} & \multicolumn{1}{c|}{{4.08 $\%$}} & \multicolumn{1}{c|}{-} \\ \hline
\multicolumn{1}{|c|}{XXL} & \multicolumn{1}{c|}{2.05$\%$} & \multicolumn{1}{c|}{12.44$\%$} & \multicolumn{1}{c|}{0.75$\%$} \\ \hline
\end{tabular}
}}
\caption{Difference in the mean expected cost between the last 10 generations of the DE. The difference is giving as a percentage with respect to the value of the mean expected cost of the last generation. We give this difference for all the 6 different sizes of the DPO problem and the 3 different ansatz circuits considered in Section~\ref{sec:PortfolioOptimization}.}
\label{tab:ConvergenceDiffEv}
\end{table}

\begin{figure}[p]
    \centering
    \includegraphics[width=.475\textwidth]{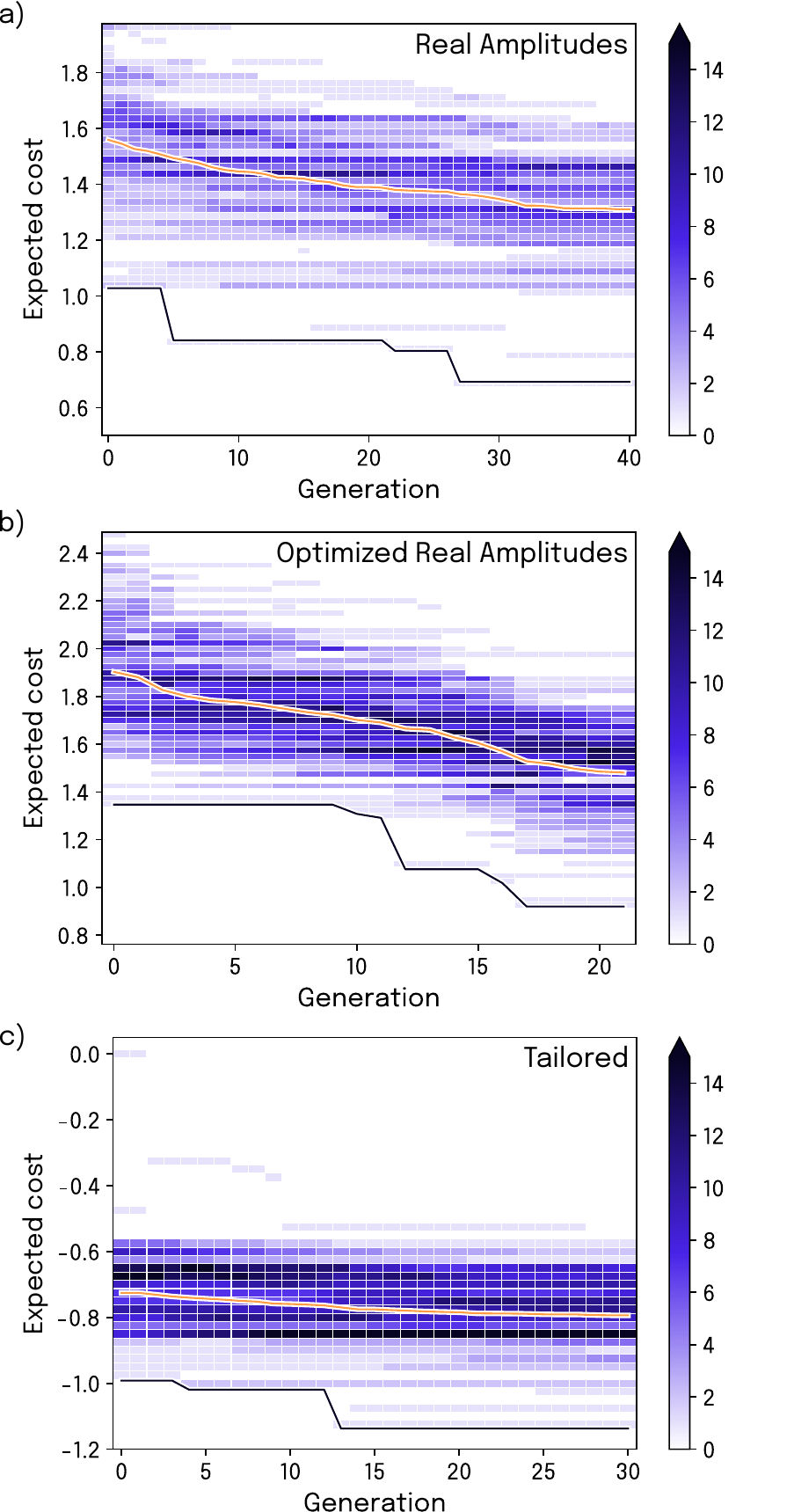}
    \vfill
    \caption{Convergence of different ansatz circuits in optimization Using the DE algorithm. Panels a), b), and c) show the convergence for the RA, ORA, and tailored ansatz circuits, respectively. For each DE generation (step) indicated on the horizontal axis, the vertical axis shows the expected optimization cost obtained using the estimator primitive (see the VQE optimizer scheme in Section~\ref{sec:VQE}). The expected cost is discretized into intervals of $0.025$. The color intensity represents the frequency of parameter sets in a generation that result in an expected cost value within that interval. The orange line in each panel indicates the average optimization cost of the generation, while the black line represents the minimum value found in each generation.}
    \label{fig:Convergence}
\end{figure}

Figures~\ref{fig:Convergence}a-c show the convergence achieved for the XXL-sized problem using different ansatz circuits, optimized with the DE algorithm as described in this subsection. The vertical axis represents the expected cost for all individuals in a fixed generation, indicated by the horizontal axis. Specifically, the expected cost is discretized into intervals of $0.025$, and the color intensity in the figure indicates the number of parameter sets in a generation that result in an expected cost value within that interval. Note that both the range of differential generations and expected costs in the figure is different for each ansatz.

Both the RA circuit (Fig.~\ref{fig:Convergence}a) and the tailored circuit (Fig.~\ref{fig:Convergence}c) converge within the specified number of generations for the XXL problem, as indicated by the mean value of the expected cost (orange lines) stabilizing within a $\leq2.5\%$ difference over the last 10 optimization steps.

In contrast, the ORA ansatz (Fig.~\ref{fig:Convergence}b) fails to converge within the given number of generations. This is likely due to the significantly larger number of parameters that need to be optimized in the ORA case (see Fig.~\ref{fig:depth_parameters}b).

Table~\ref{tab:ConvergenceDiffEv} details the convergence behavior for different problem sizes and ansatz circuits.


\section{Additional Cyclic Ansatz Example}
\label{app:AnsatzesSpec:Cyclic}
\begin{figure}[t]
    \centering
    \includegraphics[width=.475\textwidth]{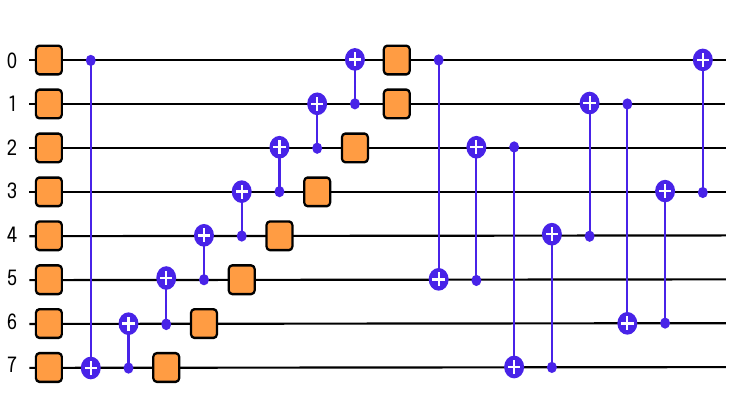}
    \caption{Scheme of cyclic ansatz for 8 qubits (This figure is for explanation purposes only, this particular circuit was not run through the development of the present paper). Each horizontal rule represents a qubit throught the execution time. The orange boxes are $R_Y$ gates and the blue circles connected to vertical lines are $\text{C-NOT}$ gates.}
    \label{fig:CyclicAnsatz8}
\end{figure}

As explained in Section~\ref{sec:cyclicCG}, the cyclic ansatz originates from a methodical assembling, highly dependent of the number of qubits selected. Typically, it exhibits some stair-shaped behavior, as showed in Fig.~\ref{fig:CyclicAnsatz8}, where we represent the cyclic ansatz for a 8 qubit problem (for example with $N_r = 2$, $N_a=2$, $N_t=2$). However, in some specific cases, like in Fig.~\ref{fig:CyclicAnsatz}, this pattern is not present due to the mathematical description of the method and the chosen range for the second block $d = 3$.

\section{Classical and Quantum computational details}

\label{app:computational_details}
All classical calculations presented in this work, including classical benchmarks and VQE assistance, were obtained using a workstation with the following specifications: 32 GB RAM, Intel Core\textsuperscript{TM} i9-13900K CPU, and NVIDIA RTX\textsuperscript{TM} A4000 GPU. 

Quantum executions were conducted on the IBM Torino QPU from September 2 to December 26, 2024. During this period, the IBM Torino QPU underwent numerous upgrades. Notably, significant enhancements were made to the Qiskit Runtime client, resulting in a substantial speedup in circuit evaluation~\cite{ibm_QDC24}. For example, we have experience a notably reduction in evaluating $\approx 3000$ circuits (involving 112 qubits and depth $~300$): initially, it took approximately $8$ hours to evaluate this QPU workload, whereas the same computation took less than $6$ hours in the last evaluations.

For the quantum computations performed with the estimator primitive~\cite{qiskit_estimator} we used $2.5 \times 10^3$ shots for the XS size and $25 \times 10^3$ shots for larger sizes. We executed the sampler primitive~\cite{qiskit_sampler} with $10^4$ shots for the XS size and $10^5$ shots for larger sizes.


\section{Data preparation}
\label{app:Data}
For this work, we extracted seven asset data between 1st January 2023 and 1st August 2023. The assets have been selected represent different asset categories: capital assets, foreign exchange asset, public bond assets, private assets, real state assets, and commodity assets. \\

\begin{figure}[t]
    \centering
    \includegraphics[width=.475\textwidth]{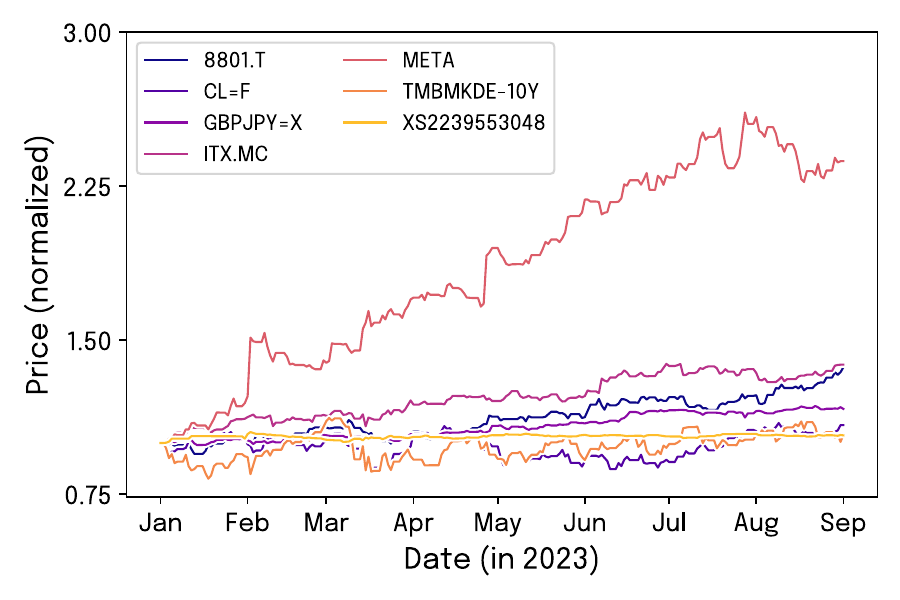}
    \caption{Daily evolution of the closing price of the seven selected asset to work with the sizes described in Table.~\ref{tab:Sizes}.}
    \label{fig:Asset_data_set1}
\end{figure}

Figure~\ref{fig:Asset_data_set1} shows the closing price evolution of these assets. In the legend of the figure we give the tickers for each asset used in our work. For the XS, S, and M sizes we use less than seven assets. In those cases we use the first $N_a$ assets in alphabetic order.


\subsubsection*{Covariance}

In Eq.~\eqref{eq:risk}, we introduce the covariance matrix $\Sigma_t$ at the investment time $t$, \textit{i.e.}, the covariances between the assets in the portfolio. To calculate $\Sigma_t$, we use the \textit{daily} logarithmic return, denoted as $\mu_{s,a}$, for each asset $a$ during the time $s$ between consecutive investments (not to be confused with $\mu_{t,a}$ in Eq.~\eqref{eq:mu}). For this work we set the time between investments to $\Delta t = 30$ days, and we calculate the daily logarithmic return as $\mu_{s,a} = \log(P_{s+1,a}/P_{s,a})$ where $P_{s,a}$ represents the closing price of asset $a$ on day $s$ (as illustrated in Fig.~\ref{fig:Asset_data_set1}). The covariance $\Sigma_t(a,b)$ between assets $a$ and $b$ at time $t$ is then calculated as follows:
\begin{align}
   &\Sigma_t(a,b) =\nonumber \\
   &\frac{1}{\Delta t - 1} \sum_{s = t\cdot \Delta t - 1}^{t\cdot \Delta t} (\mu_{s,a} - \bar{\mu}_{t,a}) (\mu_{s,b} - \bar{\mu}_{t,b}),
\end{align}
where $\bar{\mu}_{t,a}$ is the average daily logarithmic return of asset $a$ over the same interval.



\section{Details on classical Benchmarks}
\label{app:Benchmarks}

The solvers Gurobi, DOCPLEX, and GEKKO utilized in this study were based on their free versions with default configurations. For the XL and XXL experiment sizes, the execution time for DOCPLEX and Gurobi was limited to 15 minutes; for undefined time limit the models failed to converge and, after this time threshold, the optimizer does not find lower (better) optimization cost values. It is important to note that GEKKO required an initial starting point for optimization, and the results were highly dependent on this state. In this case we found that, among all possible trivial initial states, using $x_{t,a,r} = 1$ for all $t$, $a$, and $r$ give the bests results.


\section{Details on Tensor Networks Calculations}
\label{app:DetailsOnTN}

\begin{figure}[t]
\centering
\includegraphics[width=.475\textwidth]{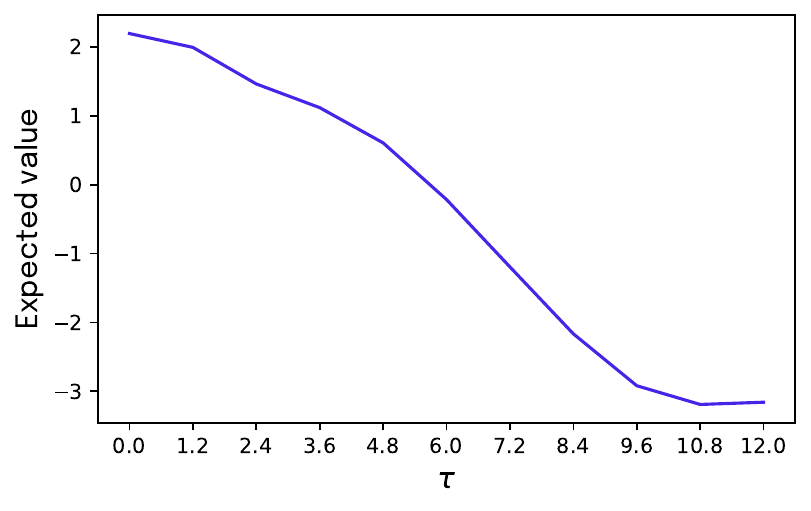}
\caption{Convergence of the SAE between times 0 and $T = 12.0$ for the XXL size, where measurements were taken at 11 equally spaced intervals.}
\label{fig:EvolutionTNXXL}
\end{figure}


For the quantum-inspired SAE, we have simulated the adiabatic temporal evolution of a thermalized Hamiltonian $H(\tau)$ as described in Eq.
(4) of Ref.~\cite{zahedinejad2017combinatorialoptimizationgatemodel}. In the reference, the evolution time parameter, $T$, is introduced, which controls the rate at which the Hamiltonian evolves. The values of $T$ chosen for the different sizes are described in Table~\ref{tab:sizeTvalues}.

\begin{table}[t]
\scalebox{0.9}{
{\renewcommand{\arraystretch}{1.25}
\centering
\begin{tabular}{|c|c|c|c|c|c|c|} \hline 
\textbf{Size} & XS & S & M & L & XL & XXL \\
\hline 
\textbf{T} & 7.0 & 19.5 & 20.0 & 7.0 & 11.0 & 12.0 \\
\hline
\end{tabular}}}
\caption{Evolution time parameter $T$ values used in the SAE for different DPO problem sizes.}
\label{tab:sizeTvalues}
\end{table}

The workflow of the algorithm involves first applying a sequence of quantum gates that corresponds to the simulation of the adiabatic process. After this, the SAE is approximated as a Matrix Product State (MPS). The sampling is then performed using the same number of shots as those used for each system size in the VQE algorithm, ensuring consistency in the comparison. 

To visualize the adiabatic evolution process, in Fig.~\ref{fig:EvolutionTNXXL}, we show the evolution of the expected value of the Hamiltonian for the XXL size. As expected, there is a gradual decrease in the expected value, which reflects that the system is approaching the ground state of the final Hamiltonian, corresponding to the optimal solution of the given problem. The process begins with an expected value of $2.20$, which corresponds to the offset value (as it represents the state of maximum entropy) and ends at $-3.20$. According to the minimum value found by Gurobi (see \ref{tab:clasicalcostbenchmarks}), the final value approaches this with a relative error of approximately $0.25$. It is important to note that this is the expected value, not the minimum value of the algorithm.

\bibliography{References}
\end{document}